\documentclass[prd,a4paper,nofootinbib,final,twocolumn,floatfix]{revtex4}
\usepackage{amsfonts}
\usepackage{amsmath}
\usepackage{amssymb}
\usepackage[usenames,dvips]{color}
\usepackage[english]{babel}
\usepackage[latin1]{inputenc}
\usepackage{amsfonts}
\usepackage{appendix}
\usepackage{graphicx}
\usepackage{graphics,rotating}
\usepackage{dcolumn}
\usepackage{bm}
\usepackage{color}
\usepackage[usenames,dvipsnames,svgnames]{xcolor}
\usepackage[colorlinks=true,
            linkcolor=blue,
            urlcolor=black,
            citecolor=blue]{hyperref}
\usepackage{hyperref}
\usepackage[T1]{fontenc}
\usepackage{multirow}
\usepackage{float}
\usepackage{subfigure}
\usepackage{enumitem}

\begin{document}

\title{Spin and quadrupolar effects in the secular evolution of precessing
compact binaries with black hole, neutron star, gravastar, or boson star
components}
\author{Zolt\'an Keresztes$^{\dag }$, M\'arton T\'apai$^{\ddag }$,
L\'aszl\'o \'Arp\'ad Gergely$^{\star }$}
\affiliation{Institute of Physics, University of Szeged, D\'om t\'er 9, Szeged 6720,
Hungary\\
$^{\dag }${\small E-mail: zkeresztes@titan.physx.u-szeged.hu\quad }$^{\ddag
} ${\small E-mail: tapai@titan.physx.u-szeged.hu\quad }$^{\star }${\small \
E-mail: gergely@physx.u-szeged.hu }}

\begin{abstract}
We discuss precessing compact binaries on eccentric orbit with gravastar,
black hole, neutron star, or boson star components. We derive the secular
evolution equations to second post-Newtonian--order accuracy, with
leading-order spin-orbit, spin-spin, and mass quadrupole-monopole
contributions included. The emerging closed system of first-order
differential equations evolves the pairs of polar and azimuthal angles of
the spin and orbital angular momentum vectors together with the periastron
angle. In contrast with the instantaneous dynamics, the secular dynamics is
autonomous. The validity of the latter is confirmed numerically, showing
that secular evolutions look like smoothed-out instantaneous evolutions over
timescales where radiation reaction is negligible. The secular evolution of
the spin polar angles and the difference of their azimuthal angles generates
a closed subsystem, which despite the apparent singularity of spherical
polar coordinates, remains well defined through aligned configurations. We
study analytically this system for the particular cases of one spin
dominating over the other and for black hole - boson star binaries of equal
masses. In the first case known large flip-flops of the smaller spin are
reproduced, when the larger spin is almost coplanar with the orbit. We also
find new, quadrupole-induced flip-flops occurring when the neutron star with
dominant spin has a quadrupolar parameter $w_{1}\approx 3$. Finally we
analyze the evolutions of the spin angles numerically by comparing the cases
when the black hole companion is either a gravastar, another black hole or a
boson star with identical mass. We find that both the amplitude and period
of the flip-flop are the largest, when the companion is a black hole. In the
case of a boson star companion the frequency of the flip-flop increases
significantly. Further, while in the case of gravastars and black holes a
swinging-type azimuthal evolution occurs, with the spins of the components
periodically overpassing each other, their sequence is conserved when the
companion is a boson star.
\end{abstract}

\maketitle

\section{Introduction}

In the O1 and O2 runs of the Advanced LIGO Detector, the LIGO Scientific
Collaboration and the Virgo Collaboration have announced a total of 11
gravitational wave detections, ten of them produced by coalescing stellar
mass black holes \cite{LIGOcatO12}, while in one case the source was a
coalescing neutron star binary \cite{LIGO5NS}. A much larger event rate has
been seen in the more sensitive O3 run, with 39 compact binary mergers
identified during its first half \cite{O3a}, among them other plausible
neutron star mergers and neutron star - black hole mergers. The spin of the
merging black holes was identified with high significance in two cases
during the O1--O2 runs \cite{LIGOcatO12,LIGOspin}, while in the first half
of the O3 run nine events were identified with the effective spin parameter
nonvanishing in the $90\%$ symmetric credible interval \cite{O3a}. The
accurate description of both the orbital and spin dynamics of compact binary
systems is important for gravitational wave source modeling; however, the
imprint of spin effects also occurs in radio astronomy. From the analysis of
the Very Long Baseline Interferometry (VLBI) radio data of a binary spanning
over 18 years the spin precession of the dominant supermassive black hole
could be identified \cite{VLBIKun}.

Whenever the black hole spins $\mathbf{S}_{\mathbf{1}}$, $\mathbf{S}_{%
\mathbf{2}}$ and the Newtonian orbital angular momentum $\mathbf{L}_{\mathbf{%
N}}$ of the binary are not aligned, they undergo precession \cite%
{barkeroconnell2,barkeroconnell1,barkeroconnell}. The total angular momentum 
$\mathbf{J}=\mathbf{L}+\mathbf{S}_{\mathbf{1}}+\mathbf{S}_{\mathbf{2}}$
composed of the total orbital angular momentum $\mathbf{L}$ and the
individual spins is conserved up to the second post-Newtonian (2PN) order 
\cite{KWW,KIDDER52,WILL,WW}, radiative dissipation appearing at 2.5PN orders 
\cite{Peters}. Both conservative and dissipative contributions to dynamics
arising from leading 1.5 PN-order spin-orbit (SO) coupling have been
thoroughly analyzed \cite{RS,GPV3}, similarly the 2PN\ spin-spin (SS) \cite%
{GSS,GSS2,KJ} and the 2PN mass quadrupole - mass monopole (QM) contributions 
\cite{poisson,GKQM,Racine,HLLR,VHF,BFH}. The 2PN self-interaction spin
effect in the radiative losses, representing the 2PN correction to the 1.5
PN order Lense-Thirring approximation, was first identified in Refs. \cite%
{GSS,GSS2} and explored later to derive the respective contributions to the
accumulated orbital phase \cite{MVG}. The dynamics of compact binary systems
has been analyzed to 4PN-order accuracy either in the nonspinning case \cite%
{4PNb,4PNa}, or in a conservative approach at the Hamiltonian level \cite%
{LS4PN}. Hamiltonian methods applied for compact binaries generated several
notable results; see Ref. \cite{HamPN} and references herein.

Compact binary dynamics in the inspiral regime exhibit three distinct
timescales. The shortest is the radial timescale, defined by the period of
the orbital motion of the reduced mass. Under the precessional timescale the
orbital angular momentum $\mathbf{L}_{\mathbf{N}}$ and the spins $\mathbf{S}%
_{\mathbf{1}}$ and $\mathbf{S}_{\mathbf{2}}$ undergo a full rotation about
their precession axis. Over the gravitational radiation reaction timescale
the effects of gravitational dissipation become noticeable. Averaging the
dynamics over any of these timescales may be rewarding. When
precession-related effects are targeted, averaging over a radial period will
remove insignificant instantaneous effects, but keep the dominant
precessional evolution. Several interesting spin-related evolutions were
identified by this method. When the orbital angular momentum nearly cancels
the total spin the orbital plane changes significantly during a relatively
short-lived transitional precession period \cite{ACST}. The direction of
dominant spin relative to the total angular momentum can change
significantly over the radiation reaction timescale in binaries where the
components have significantly different mass \cite{spinflip}. This spin flip
may explain the formation of X-shape radio galaxies \cite{xshaped}. By an
additional averaging over the precession timescale the evolution of the
magnitude of total angular momentum over the radiation reaction timescale
was investigated in Refs. \cite{kesden1,kesden2}.

When the dominant spin vector is approximately perpendicular to $\mathbf{L}_{%
\mathbf{N}}$ and the smaller spin is closely aligned with it, the smaller
spin slowly evolves to be antialigned with $\mathbf{L}_{\mathbf{N}}$, then
periodically changes back and forth on a timescale shorter than the
gravitational radiation reaction timescale \cite{flipflop}. This effect was
investigated in a wider parameter range for binaries moving on circular
orbits \cite{flipflop1,flipflop2}. This spin flip-flop effect was first
found qualitatively in Ref. \cite{majar} as a harmonic "wooble" in the polar
angle of the spin, which evolves "from pole to pole". Recently, a parameter
range has been identified where the flip-flop happens on relatively short
timescales, dubbed as wide precession \cite{Gerosa2019}. Then over the
period during which the magnitude of the total spin changes from its minimum
to its maximum and back to the minimum value, one of the two spins evolves
from complete alignment with $\mathbf{L}_{\mathbf{N}}$ to complete
antialignment.

Kidder has derived the orbit-averaged (secular) spin-precession equations 
\cite{KIDDER52} for circular orbits, with SO and SS contributions included,
but QM contributions omitted. The QM couplings were included in the
discussion of angular evolutions in black hole binaries (where the
quadrupole parameters are $w_{i}=1$) by Racine \cite{Racine}, who presented
a new constant of motion $\lambda $ of the orbit-averaged dynamics (or
equivalently $\xi =\lambda L_{N},$ the last factor representing the
magnitude of $\mathbf{L}_{\mathbf{N}}$).

Racine also solved analytically the averaged equations for equal masses and
derived approximate analytical solutions in the unequal mass case. His
analytical solution has been generalized for arbitrary masses and spins (but
still $w_{i}=1$ and circular orbits) in Ref. \cite{KGSBS}, which identified
three distinct regimes in the orbit-averaged precessional dynamics:
librations about the configurations of the two spin projections to the
orbital plane either aligned or antialigned and a "circulating"
configuration, when one of the spins precesses much faster.

Because of the recent discovery of gravitational waves from mergers of
neutron stars \cite{LIGO5NS,2NS}, the interest in their internal structure
and equation of state \cite{RPZB}, implying a better constrained range of
the parameter $w_{i}$ has been revitalized. In this work we leave the
quadrupole parameter unspecified, including neutrons stars with $w_{i}\in $ $%
\left( 2,14\right) $ \cite{NSw}, \cite{NSUrbanecw} and also other exotic
compact objects, like boson stars with $w_{i}\in \left( 10,150\right) $ \cite%
{BosonStarw} or gravastars with $w_{i}\in \left( -0.8,1\right) $ \cite%
{GravaStarw} as compact binary components. Allowing for $w_{i}\neq 1$
complicates the dynamics, as for example the quantity conserved for black
hole binaries, identified by Racine, becomes dynamical, unless $w_{i}=1$.

Although gravitational radiation tends to circularize the orbit of the
binary \cite{Peters}, significant eccentricity can be preserved at the end
of the inspiral. This happens for binaries in either dense galactic nuclei 
\cite{ecc01,ecc02} or within accretion disks \cite{ecc03,ecc04}.
Furthermore, because of the Kozai mechanism, the relativistic orbital
resonances in hierarchical triples can also retain eccentricity \cite%
{ecc05,ecc06,ecc07,ecc08}. The interaction between supermassive black hole
binaries and\ their star populations results in significant eccentricity
toward the end of the inspiral too \cite{ecc09,ecc10}. Hence allowing for
eccentricity in the dynamics may be rewarding.

The instantaneous dynamics (including SO, SS and QM effects, also
eccentricity) in terms of dimensionless variables was discussed in Refs. 
\cite{Inspiral1,Inspiral2,chameleon}, based on earlier works on binary
dynamics of Refs. \cite%
{barkeroconnell2,barkeroconnell1,barkeroconnell,KWW,damourderuelle,blanchet,OBKPSSZ,KMG}%
. The 2PN conservative dynamics of compact binary systems was given by Eqs.
(36)-(42) of Ref. \cite{chameleon} in terms of dimensionless osculating
orbital elements $\mathfrak{l}_{r}$, $e_{r}$, $\psi _{p}$, $\alpha $, and $%
\phi _{n}$; spin polar and azimuthal angles $\kappa _{i}$ and $\zeta _{i}$ ($%
i=1,2$); and true anomaly parameter $\chi _{p}$. The time evolution of $\chi
_{p}$ is governed by Eq. (43) of Ref. \cite{chameleon}. The polar angles $%
\kappa _{i}$ of the spin vectors are measured from $\mathbf{L}_{\mathbf{N}}$%
. The azimuthal angles $\zeta _{i}$ are measured from the Laplace-Runge-Lenz
vector $\mathbf{A}_{\mathbf{N}}$ in the plane perpendicular to $\mathbf{L}_{%
\mathbf{N}}$. The argument of the periastron, $\psi _{p}$ is defined by $%
\psi _{p}=\arccos \left( \mathbf{\hat{l}}\cdot \mathbf{\hat{A}}_{\mathbf{N}%
}\right) $, with $\mathbf{\hat{l}}=\mathbf{\hat{J}}\times \mathbf{\hat{L}}_{%
\mathbf{N}}$, where $\mathbf{\hat{J}}$ is the direction of the total angular
momentum which is conserved in the 2PN dynamics. The inclination $\alpha $
is the polar angle of $\mathbf{\hat{L}}_{\mathbf{N}}$ measured from $\mathbf{%
\hat{J}}$. The last angle is the longitude of the ascending node $-\phi _{n}$%
, spanned by the inertial axis $\mathbf{\hat{x}}$ (arbitrarily chosen in the
plane perpendicular to $\mathbf{\hat{J}}$) and $\mathbf{\hat{l}}$. This
angle is related to the azimuthal angle $\left( \pi /2-\phi _{n}\right) $ of 
$\mathbf{\hat{L}}_{\mathbf{N}}$, measured from $\mathbf{\hat{x}}$ in the
plane perpendicular to $\mathbf{\hat{J}}$. (The angles $\chi _{p}$, $\kappa
_{i}$, $\psi _{i}=\zeta _{i}-\psi _{p}$, $\psi _{p}$, $\alpha $, and $\phi
_{n}$ are shown in Fig. 2 in Ref. \cite{Inspiral1}.)

The dimensionless orbital angular momentum 
\begin{equation}
\mathfrak{l}_{r}=\frac{cL_{N}}{Gm\mu }~
\end{equation}%
and the eccentricity%
\begin{equation}
e_{r}=\frac{A_{N}}{Gm\mu }~
\end{equation}%
characterize the osculating ellipse of the orbit; hence, they are shape
variables. The total and reduced masses of the binary are denoted as $%
m=m_{1}+m_{2}$ and $\mu =m_{1}m_{2}/m$, respectively. We also employ the
mass ratio $\nu =m_{2}/m_{1}\leq 1$ and the symmetric mass ratio $\eta =\mu
/m$. The gravitational constant, the speed of light, and the magnitude of $%
\mathbf{A}_{\mathbf{N}}$ are denoted by $G$, $c$, and $A_{N}$ respectively.
The magnitudes of the spins are characterized by the dimensionless spin
parameters $\chi _{i}=cS_{i}/Gm_{i}^{2}$ $\left( i=1,2\right) $. The dot
will denote the derivative with respect to the dimensionless time variable $%
\mathfrak{t}=tc^{3}/Gm$ (with time $t$) introduced in Ref. \cite{chameleon}.
The PN parameter is defined as $\varepsilon =Gm/c^{2}r$.

In this paper, we will investigate precessing compact binary systems on
eccentric orbit subject to bound motion, first establishing the 2PN secular
dynamics in terms of the above-mentioned dimensionless variables, then
analyzing the spin evolutions with the methods of dynamical systems, with
SO, SS and QM\ contributions included. The PN equations of motion depend on
the choice of a spin supplementary condition (SSC)\footnote{%
A comparison of the three widely used SSCs has been presented in Ref.\textbf{%
\ }\cite{mikoczissc}, proving the SSC dependence of the radiative multipole
moments.}. We employ the Newton-Wigner-Price \cite{NWP1,NWP2} SSC, similarly
as in Ref. \cite{chameleon}. This system is valid for eccentric orbits and
for binaries composed of either black holes, neutron stars or other exotic
compact objects (boson stars or gravastars).

The instantaneous dynamics described in Ref. \cite{chameleon} is averaged
over a suitably defined radial period in Section \ref{secdyn}, obtaining the
secular precessing compact binary dynamics in terms of the dimensionless
osculating orbital elements $\mathfrak{l}_{r}$, $e_{r}$, $\psi _{p}$, $%
\alpha $, $\phi _{n}$ and spin angles $\kappa _{i}$, $\zeta _{i}$ ($i=1,2$).
These equations contain PN, SO, SS, QM and 2PN contributions. These are
generalized Lagrangian planetary equations, which become singular for
vanishing $e_{r}$, nevertheless the singularity can be eliminated by a
transformation of variables \cite{LagrangePlanetarytrafo}. For completeness,
we also give in this section the secular precession angular velocities and
constraints relating the variables.

For the purpose of averaging, the PN expansion of the radial period is
required. For clarity of presentation we deferred the tedious but
straightforward bulk of calculations leading to it to the lengthy Appendix %
\ref{appendix1}. There, first, we derive the radial period in terms of the
variables evaluated at the periastron (characterized by the value of the
true anomaly parameter $\chi _{p}=0$). The $\chi _{p}$ dependence of the
shape variables is also derived by integrating the corresponding first-order
system given in Ref. \cite{chameleon}. Then, we compute their time-averaged
values $\bar{\mathfrak{l}}_{r}$, $\bar{e}_{r}$ over the radial period to 2PN
accuracy, with the inclusion of all spin and mass quadrupole effects to this
order. Next, we express the shape variables evaluated in the periastron in
terms of the corresponding averaged quantities. With this, we can write the
radial period as a PN expansion expressed in terms of averaged quantities.
We also give there a similar expansion in terms of the chosen variables of
the PN parameter.

In Sec. \ref{limits}, we analyze the role of eccentricity in the secular
evolution by comparing low-eccentricity and medium-eccentricity evolutions
for the values of the PN parameter $0.01$ and $0.0005$. We also prove that
the secular dynamics follows closely the instantaneous dynamics given in
Ref. \cite{chameleon} over the conservative timescale.

The secular evolution of the spin angles generates a closed subsystem of
three variables, discussed in Sec. \ref{closed}. Despite the apparent
singularity of spherical polar coordinates, we prove in Appendix \ref{reg}
that the system remains well defined through aligned configurations.

In Sec. \ref{particular}, we study analytically and numerically this closed
subsystem for the particular case of one spin dominating over the other,
concentrating on the flip-flop effect of the polar angle of the smaller
spin. We identify a diamond-shaped region in the parameter plane span by the
dominant spin polar angle and quadrupolar parameter. Along the horizontal
axis of the diamond the known flip-flop effect for black hole binaries is
reproduced; however, the vertical axis signifies mass quadrupole-induced
flip-flops occurring for neutron stars with a particular quadrupolar
parameter.

In Sec. \ref{boson}, we study the closed subsystem for the particular case
of black hole - boson star binaries. We investigate the spin angle dynamics
both analytically and numerically. We also compare it to typical evolutions
in black hole - black hole and black hole - gravastar binaries, pointing out
the differences.

In Sec. \ref{concludingr}, we present the conclusions.

\section{Secular conservative dynamics of precessing compact binaries \label%
{secdyn}}

In this section we present the orbital averaged evolution equations of the
dimensionless osculating orbital elements $\mathfrak{l}_{r}$, $e_{r}$, $\psi
_{p}$, $\alpha $, and $\phi _{n}$ and spin angles $\kappa _{i}$, $\zeta _{i}$
($i=1,2$) at 2PN accuracy, with spin and mass quadrupole effects included
from the instantaneous evolutions derived in Ref. \cite{chameleon}.

\subsection{Averaging method\label{averaging}}

For bounded motion the separation $r$ between the binary components can be
parametrized similarly to a Keplerian orbit \cite{chameleon}:%
\begin{equation}
\frac{c^{2}r}{Gm}=\frac{\mathfrak{l}_{r}^{2}}{1+e_{r}\cos \chi _{p}}~,
\end{equation}%
where $\chi _{p}$ is the true anomaly. However, unlike a Keplerian orbit,
both shape variables $\mathfrak{l}_{r}$ and $e_{r}$ are time dependent.

The dimensionless period $\mathfrak{T}$ is defined as the change in the
dimensionless time $\mathfrak{t}\equiv tc^{3}/Gm$ during the evolution of
the true anomaly over $\chi _{p}\in \left[ 0,2\pi \right] $ as 
\begin{equation}
\mathfrak{T}\equiv \int_{0}^{\mathfrak{T}}d\mathfrak{t}=\int_{0}^{2\pi }%
\frac{d\chi _{p}}{\dot{\chi}_{p}}~.  \label{perioddef}
\end{equation}%
The average $\bar{f}$ of any quantity $f\left( \mathfrak{t}\right) $ with
respect to $\mathfrak{t}$ is introduced by%
\begin{equation}
\mathfrak{T}\bar{f}=\int_{0}^{\mathfrak{T}}f\left( \mathfrak{t}\right) d%
\mathfrak{t}=\int_{0}^{2\pi }\frac{f\left( \chi _{p}\right) }{\dot{\chi}_{p}}%
d\chi _{p}~.  \label{averdef}
\end{equation}%
Then, as described in Appendix \ref{appendix1} the period can be rewritten
in terms of the orbital averaged shape variables as 
\begin{eqnarray}
\mathfrak{T} &=&\mathfrak{\tilde{T}}\left( 1+\frac{1}{\mathfrak{\bar{l}}%
_{r}^{2}}\tilde{\tau}_{PN}+\frac{1}{\mathfrak{\bar{l}}_{r}^{3}}\tilde{\tau}%
_{SO}+\frac{1}{\mathfrak{\bar{l}}_{r}^{4}}\tilde{\tau}_{QM}\right.  \notag \\
&&\left. +\frac{1}{\mathfrak{\bar{l}}_{r}^{4}}\tilde{\tau}_{SS}+\frac{1}{%
\mathfrak{\bar{l}}_{r}^{4}}\tilde{\tau}_{2PN}\right) ~,  \label{periodaver}
\end{eqnarray}%
with the expressions $\mathfrak{\tilde{T}}$, $\tilde{\tau}_{PN}$,~$\tilde{%
\tau}_{SO}$, $\tilde{\tau}_{QM}$,$~\tilde{\tau}_{SS}$, and$~\tilde{\tau}%
_{2PN}$ also given there. Here, $1/\mathfrak{l}_{r}^{2}$ stands for one
relative PN\ order, as indicated by Eq. (8) of Ref. \cite{chameleon}.

\subsection{Shape variables\label{shape}}

Employing the averaging method described in the previous section, long but
straightforward calculations lead to the secular evolutions of the
dimensionless orbital angular momentum $\mathfrak{l}_{r}$ and orbital
eccentricity $e_{r}$ as 
\begin{equation}
\mathfrak{\bar{\dot{l}}}_{r}=\mathfrak{\bar{\dot{l}}}_{r}^{PN}=\mathfrak{%
\bar{\dot{l}}}_{r}^{SO}=\mathfrak{\bar{\dot{l}}}_{r}^{SS}=\mathfrak{\bar{%
\dot{l}}}_{r}^{QM}=\mathfrak{\bar{\dot{l}}}_{r}^{2PN}=0~,
\end{equation}%
\begin{equation}
\bar{\dot{e}}_{r}=\bar{\dot{e}}_{r}^{PN}=\bar{\dot{e}}_{r}^{SO}=\bar{\dot{e}}%
_{r}^{SS}=\bar{\dot{e}}_{r}^{QM}=\bar{\dot{e}}_{r}^{2PN}=0~.
\end{equation}%
As expected, the average shape of the orbit does not change when dissipation
by gravitational waves is neglected.

\subsection{Euler angles\label{Euler}}

These evolutions are nontrivial, as discussed below.

\subsubsection{Inclination $\protect\alpha $}

The secular evolution of the inclination $\alpha =\arccos \left( \mathbf{%
\hat{J}}\cdot \mathbf{\hat{L}}_{\mathbf{N}}\right) $ emerges as the sum of
the contributions:%
\begin{equation}
\bar{\dot{\alpha}}^{PN}=0~,
\end{equation}%
\begin{eqnarray}
\bar{\dot{\alpha}}^{SO} &=&\frac{\eta \pi }{\mathfrak{T}~\mathfrak{\bar{l}}%
_{r}^{3}}\sum_{k=1}^{2}\left( 4\nu ^{2k-3}+3\right)  \notag \\
&&\times \chi _{k}\sin \kappa _{k}\cos \left( \psi _{p}+\zeta _{k}\right) ~,
\label{alphaSO}
\end{eqnarray}%
\begin{eqnarray}
\bar{\dot{\alpha}}^{SS} &=&-\frac{3\eta \pi }{\mathfrak{T~\bar{l}}_{r}^{4}~}%
\chi _{1}\chi _{2}  \notag \\
&&\times \left[ \sin \kappa _{1}\cos \kappa _{2}\cos \left( \psi _{p}+\zeta
_{1}\right) \right.  \notag \\
&&\left. +\cos \kappa _{1}\sin \kappa _{2}\cos \left( \psi _{p}+\zeta
_{2}\right) \right] ~,  \label{alphaSS}
\end{eqnarray}%
\begin{eqnarray}
\bar{\dot{\alpha}}^{QM} &=&-\frac{3\eta \pi }{2\mathfrak{T~\bar{l}}_{r}^{4}~}%
\sum_{k=1}^{2}\nu ^{2k-3}w_{k}\chi _{k}^{2}  \notag \\
&&\times \sin 2\kappa _{k}\cos \left( \psi _{p}+\zeta _{k}\right) ~.
\label{alphaQM}
\end{eqnarray}%
\begin{equation}
\bar{\dot{\alpha}}^{2PN}=0~.
\end{equation}%
As expected, the inclination only changes due to spin and quadrupolar
effects.

\subsubsection{Longitude of the ascending node $-\protect\phi _{n}$}

The longitude of the ascending node $-\phi _{n}$ is subtended by the
inertial axis $\mathbf{\hat{x}}$ and the ascending node $\mathbf{\hat{l}}=%
\mathbf{\hat{L}}_{\mathbf{N}}\times \mathbf{\hat{J}}$. It has the following
contributions to its secular evolution:%
\begin{equation}
\bar{\dot{\phi}}_{n}^{PN}=0~,
\end{equation}%
\begin{eqnarray}
\bar{\dot{\phi}}_{n}^{SO} &=&-\frac{\eta \pi }{\mathfrak{T~\bar{l}}%
_{r}^{3}\sin \alpha }\sum_{k=1}^{2}\left( 4\nu ^{2k-3}+3\right)  \notag \\
&&\times \chi _{k}\sin \kappa _{k}\sin \left( \psi _{p}+\zeta _{k}\right) ~,
\end{eqnarray}%
\begin{eqnarray}
\bar{\dot{\phi}}_{n}^{SS} &=&\frac{3\eta \pi }{\mathfrak{T}~\mathfrak{\bar{l}%
}_{r}^{4}\sin \alpha }\chi _{1}\chi _{2}  \notag \\
&&\times \left[ \sin \kappa _{1}\cos \kappa _{2}\sin \left( \psi _{p}+\zeta
_{1}\right) \right.  \notag \\
&&\left. +\cos \kappa _{1}\sin \kappa _{2}\sin \left( \psi _{p}+\zeta
_{2}\right) \right] ~,
\end{eqnarray}%
\begin{eqnarray}
\bar{\dot{\phi}}_{n}^{QM} &=&\frac{3\eta \pi }{2\mathfrak{T}~\mathfrak{\bar{l%
}}_{r}^{4}\sin \alpha }\sum_{k=1}^{2}\nu ^{2k-3}w_{k}\chi _{k}^{2}  \notag \\
&&\times \sin 2\kappa _{k}\sin \left( \psi _{p}+\zeta _{k}\right) ~,
\end{eqnarray}%
\begin{equation}
\bar{\dot{\phi}}_{n}^{2PN}=0~.
\end{equation}%
Again, only spin and quadrupolar effects contribute.

\subsubsection{Argument of the periastron $\protect\psi _{p}$}

The secular evolution of $\psi _{p}$, the angle between the node line ($%
\mathbf{\hat{l}}$, perpendicular to both $\mathbf{L}_{\mathbf{N}}$ and $%
\mathbf{J}$) and the periastron ($\mathbf{\hat{A}}_{\mathbf{N}}$), is the
sum of:%
\begin{equation}
\bar{\dot{\psi _{p}}}^{PN}=\frac{6\pi }{\mathfrak{T}~\mathfrak{\bar{l}}%
_{r}^{2}}~,  \label{psipPN}
\end{equation}%
\begin{eqnarray}
\bar{\dot{\psi _{p}}}^{SO} &=&-\frac{\eta \pi }{\mathfrak{T}~\mathfrak{\bar{l%
}}_{r}^{3}}\sum_{k=1}^{2}\left( 4\nu ^{2k-3}+3\right)  \notag \\
&&\times \chi _{k}\left[ 2\cos \kappa _{k}\right.  \notag \\
&&\left. +\cot \alpha \sin \kappa _{k}\sin \left( \psi _{p}+\zeta
_{k}\right) \right] ~,  \label{psipSO}
\end{eqnarray}%
\begin{eqnarray}
\bar{\dot{\psi _{p}}}^{SS} &=&\frac{3\eta \pi }{\mathfrak{T}~\mathfrak{\bar{l%
}}_{r}^{4}}\chi _{1}\chi _{2}  \notag \\
&&\times \left\{ \cot \alpha \left[ \sin \kappa _{1}\cos \kappa _{2}\sin
\left( \psi _{p}+\zeta _{1}\right) \right. \right.  \notag \\
&&\left. +\cos \kappa _{1}\sin \kappa _{2}\sin \left( \psi _{p}+\zeta
_{2}\right) \right] +2\cos \kappa _{1}  \notag \\
&&\left. \times \cos \kappa _{2}-\sin \kappa _{1}\sin \kappa _{2}\cos
\,\left( \zeta _{2}-\zeta _{1}\right) \right\} ~,
\end{eqnarray}

\begin{eqnarray}
\bar{\dot{\psi _{p}}}^{QM} &=&\frac{3\eta \pi }{2\mathfrak{T}~\mathfrak{\bar{%
l}}_{r}^{4}}\sum_{k=1}^{2}\nu ^{2k-3}w_{k}\chi _{k}^{2}  \notag \\
&&\times \left[ \cot \alpha \sin 2\kappa _{k}\sin \left( \psi _{p}+\zeta
_{k}\right) \right.  \notag \\
&&\left. -3\sin ^{2}\kappa _{k}+2\right] ~,
\end{eqnarray}%
\begin{equation}
\bar{\dot{\psi _{p}}}^{2PN}=\frac{3\pi }{2\mathfrak{T\bar{l}}_{r}^{4}}\left[
\allowbreak 33\bar{e}_{r}^{2}-4\eta -6\bar{e}_{r}^{2}\eta +2\right] ~.
\end{equation}%
All PN, spin and quadrupolar corrections lead to periastron precession.

\subsection{Spin angles\label{spin}}

The contributions to the secular evolutions of the spin polar angles $\kappa
_{i}$ and the azimuthal angles $\zeta _{i}$ are 
\begin{equation}
\bar{\dot{\kappa _{i}}}^{PN}=0~,  \label{kappaiPN}
\end{equation}%
\begin{eqnarray}
\bar{\dot{\kappa _{i}}}^{SO} &=&\frac{\eta \pi }{\mathfrak{T}~\mathfrak{\bar{%
l}}_{r}^{3}}  \notag \\
&&\times \left( 4\nu ^{2j-3}+3\right) \chi _{j}\sin \kappa _{j}\sin \left(
\zeta _{i}-\zeta _{j}\right) ~,
\end{eqnarray}%
\begin{eqnarray}
\bar{\dot{\kappa _{i}}}^{SS} &=&-\frac{\eta \pi }{\mathfrak{T}~\mathfrak{%
\bar{l}}_{r}^{4}}\chi _{j}\sin \kappa _{j}  \notag \\
&&\times \sin \left( \zeta _{i}-\zeta _{j}\right) \left( \mathfrak{\bar{l}}%
_{r}\nu ^{2j-3}+3\chi _{i}\cos \kappa _{i}\right) ~,  \label{kappaiSS}
\end{eqnarray}%
\begin{eqnarray}
\bar{\dot{\kappa _{i}}}^{QM} &=&-\frac{3\eta \pi }{2\mathfrak{T}~\mathfrak{%
\bar{l}}_{r}^{4}}\nu ^{2j-3}w_{j}\chi _{j}^{2}  \notag \\
&&\times \sin 2\kappa _{j}\sin \left( \zeta _{i}-\zeta _{j}\right) ~,
\end{eqnarray}%
\begin{equation}
\bar{\dot{\kappa _{i}}}^{2PN}=0~,  \label{kappai2PN}
\end{equation}%
\begin{equation}
\bar{\dot{\zeta _{i}}}^{PN}=-\bar{\dot{\psi _{p}}}^{PN}~,  \label{zetaiPN}
\end{equation}%
\begin{eqnarray}
\bar{\dot{\zeta _{i}}}^{SO} &=&\frac{\eta \pi }{\mathfrak{T}~\mathfrak{\bar{l%
}}_{r}^{3}}\left\{ \mathfrak{\bar{l}}_{r}\left( 4+3\nu ^{3-2i}\right) \right.
\notag \\
&&+3\left( 4\nu ^{2i-3}+3\right) \chi _{i}\cos \kappa _{i}+\left( 4\nu
^{2j-3}+3\right) \chi _{j}  \notag \\
&&\times \left. \left[ 2\cos \kappa _{j}+\cot \kappa _{i}\sin \kappa
_{j}\cos \left( \zeta _{i}-\zeta _{j}\right) \right] \right\} ~,
\label{zetaiSO}
\end{eqnarray}%
\begin{eqnarray}
\bar{\dot{\zeta _{i}}}^{SS} &=&-\frac{\eta \pi }{\mathfrak{T}~\mathfrak{\bar{%
l}}_{r}^{3}}\nu ^{2j-3}\chi _{j}\left[ 2\cos \kappa _{j}\right.  \notag \\
&&\left. +\cot \kappa _{i}\sin \kappa _{j}\cos \left( \zeta _{i}-\zeta
_{j}\right) \right]  \notag \\
&&-\frac{3\eta \pi }{\mathfrak{T}~\mathfrak{\bar{l}}_{r}^{4}}\chi _{i}\chi
_{j}  \notag \\
&&\times \left\{ \cot \kappa _{i}\left[ 3\sin \kappa _{i}\cos \kappa
_{j}\right. \right.  \notag \\
&&\left. +\cos \kappa _{i}\sin \kappa _{j}\cos \left( \zeta _{i}-\zeta
_{j}\right) \right]  \notag \\
&&\left. -\sin \kappa _{i}\sin \kappa _{j}\cos \left( \zeta _{i}-\zeta
_{j}\right) \right\} ~,  \label{zetaiSS}
\end{eqnarray}%
\begin{eqnarray}
\bar{\dot{\zeta _{i}}}^{QM} &=&-\frac{3\eta \pi }{\mathfrak{T}~\mathfrak{%
\bar{l}}_{r}^{3}}w_{i}\chi _{i}\cos \kappa _{i}  \notag \\
&&-\frac{3\eta \pi }{2\mathfrak{T}~\mathfrak{\bar{l}}_{r}^{4}}%
\sum_{k=1}^{2}w_{k}\nu ^{2k-3}\chi _{k}^{2}  \notag \\
&&\times \left[ 2-3\sin ^{2}\kappa _{k}\right.  \notag \\
&&\left. +\cot \kappa _{i}\sin \left( 2\kappa _{k}\right) \cos \left( \zeta
_{i}-\zeta _{k}\right) \right] ~,
\end{eqnarray}%
\begin{equation}
\bar{\dot{\zeta _{i}}}^{2PN}=-\bar{\dot{\psi _{p}}}^{2PN}~.
\end{equation}%
Here, $i\neq jm,$ and $i=1$, $2$.

It is easy to check that in the binary black hole case the dynamics indeed
provides the additional constant of motion found by Racine \cite{Racine},
which in our notation becomes%
\begin{equation}
\xi =\frac{G\mu m}{c}\left[ \left( 1+\nu ^{-1}\right) \chi _{1}\cos \kappa
_{1}+\left( 1+\nu \right) \chi _{2}\cos \kappa _{2}\right]
\end{equation}%
Its time derivative indeed vanishes as can be seen by employing Eqs. (\ref%
{kappaiPN})--(\ref{kappai2PN}) with $w_{i}=1$. However, there is no obvious
way to generalize this constant for arbitrary quadrupole parameter $%
w_{i}\neq 1$.

\subsection{Secular precession angular velocities}

The averaged precession angular velocities are calculated from Eqs.
(31)-(33) of Ref. \cite{chameleon}\footnote{%
In Eqs. (B34) of Ref. \cite{Inspiral2} the SS and QM terms have typos: the $%
1/2$ factors should be removed. We thank Krisztina Kövér for pointing this
out. Because of this, the second term of Eq. (33) of Ref. \cite{chameleon}
contains unnecessary $1/2$ factors on the rhs (but otherwise all conclusions
remain unchanged). In the present paper, these have been corrected, and both
the instantaneous and secular dynamics are represented correctly.}.%
\begin{eqnarray}
\overline{\mathbf{\omega }_{\mathbf{i}}\cdot \mathbf{\hat{A}}_{\mathbf{N}}}
&=&\frac{2\eta \pi }{\mathfrak{T}~\mathfrak{\bar{l}}_{r}^{3}}\left( \nu
^{2j-3}\chi _{j}\sin \kappa _{j}\cos \zeta _{j}\right.  \notag \\
&&\left. +3w_{i}\chi _{i}\sin \kappa _{i}\cos \zeta _{i}\right)
\end{eqnarray}%
\begin{eqnarray}
\overline{\mathbf{\omega }_{\mathbf{i}}\cdot \mathbf{\hat{Q}}_{\mathbf{N}}}
&=&\frac{2\eta \pi }{\mathfrak{T}~\mathfrak{\bar{l}}_{r}^{3}}\left( \nu
^{2j-3}\chi _{j}\sin \kappa _{j}\sin \zeta _{j}\right.  \notag \\
&&\left. +3w_{i}\chi _{i}\sin \kappa _{i}\sin \zeta _{i}\right)
\end{eqnarray}%
\begin{eqnarray}
\overline{\mathbf{\omega }_{\mathbf{i}}\cdot \mathbf{\hat{L}}_{\mathbf{N}}}
&=&\frac{\eta \pi }{\mathfrak{T}~\mathfrak{\bar{l}}_{r}^{3}}\left[ ~%
\mathfrak{\bar{l}}_{r}\left( 4+3\nu ^{3-2i}\right) \right.  \notag \\
&&\left. +2\nu ^{2j-3}\chi _{j}\cos \kappa _{j}\right]
\end{eqnarray}

It is easy to see by checking the leading-order term of $\mathfrak{T}$ that
as $\bar{e}_{r}$ goes to $1$ the precession becomes increasingly small. It
is explained by the fact that on parabolic orbits, when $\bar{e}_{r}=1$, the
motion becomes unbound, and there is no well-defined period, thus no
precession.

\subsection{Constraints}

\begin{figure*}[th]
\includegraphics[scale=0.53]{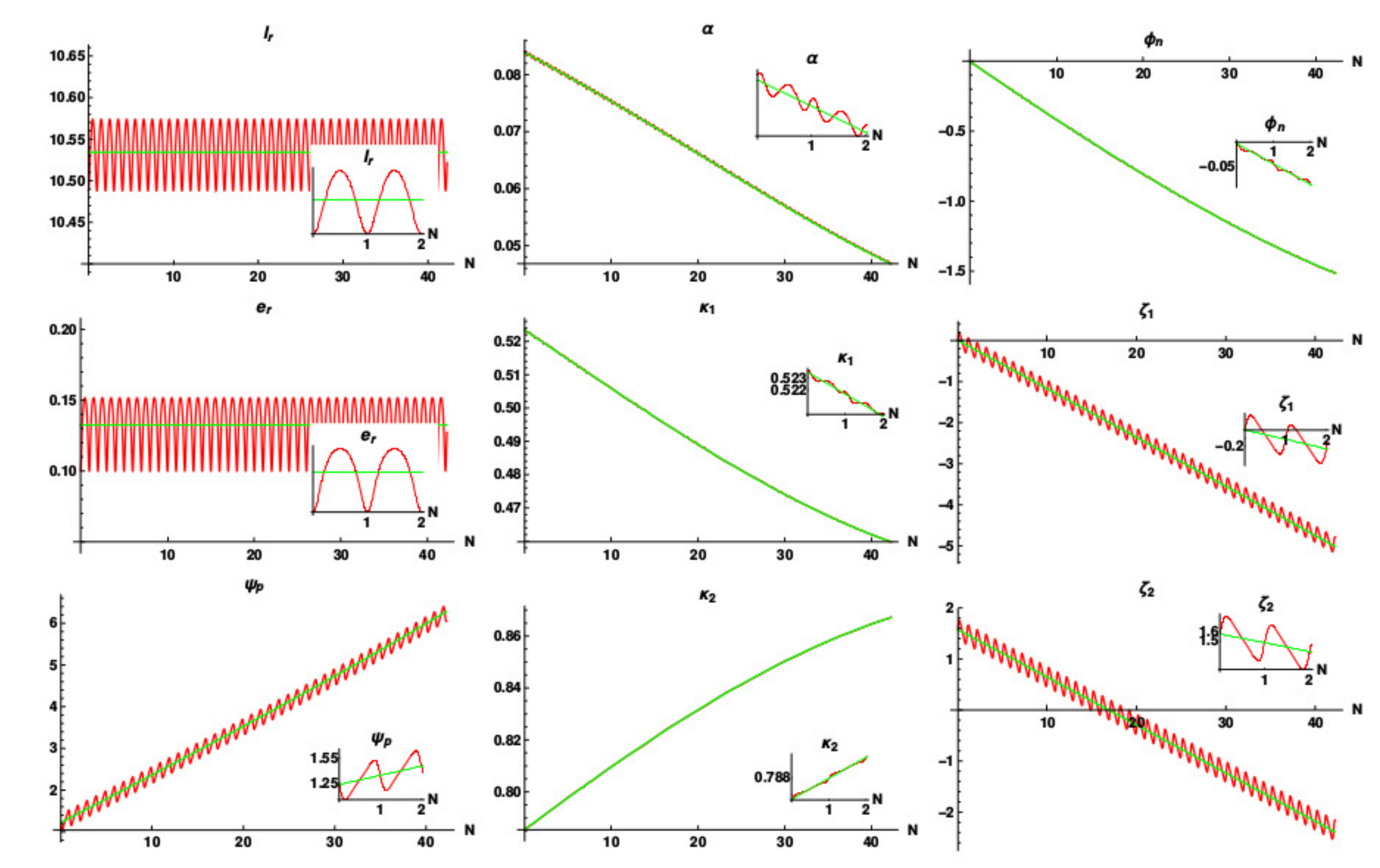} \centering
\caption{The panels represent the secular (green) and instantaneous (red)
evolutions of $\mathfrak{l}_{r}$, $e_{r}$, $\protect\alpha $, $\protect\phi %
_{n}$, $\protect\psi _{p}$, $\protect\kappa _{i}$ and $\protect\zeta _{i}$ ($%
i=1,2$) as functions of the number of orbital periods $N$. The parameters
and initial conditions for the instantaneous evolutions are: total mass {$%
m=20M_{\odot }$}, mass ratio $\protect\nu =0.5$, dimensionless spin
parameters $\protect\chi _{1}=\protect\chi _{2}=0.9982$, eccentricity $%
e_{r}\left( 0\right) =0.1$, spin polar angles $\protect\kappa _{1}\left(
0\right) =\protect\pi /6$ and $\protect\kappa _{2}\left( 0\right) =\protect%
\pi /4$, spin azimuthal angles $\protect\zeta _{1}\left( 0\right) =0$ and $%
\protect\zeta _{2}\left( 0\right) =\protect\pi /2$, longitude of the
ascending node $\protect\phi _{n}\left( 0\right) =0$ and post-Newtonian
parameter $\protect\varepsilon \left( 0\right) =0.01$. Matching initial
conditions for the secular dynamics have been chosen, as explained in the
main text. On some of the plots the two evolutions overlap at the chosen
resolution. On each panel the first two periods are also shown in order to
illustrate that the match only occurs on the longer run. }
\label{fig01}
\end{figure*}
\begin{figure*}[th]
\includegraphics[scale=0.53]{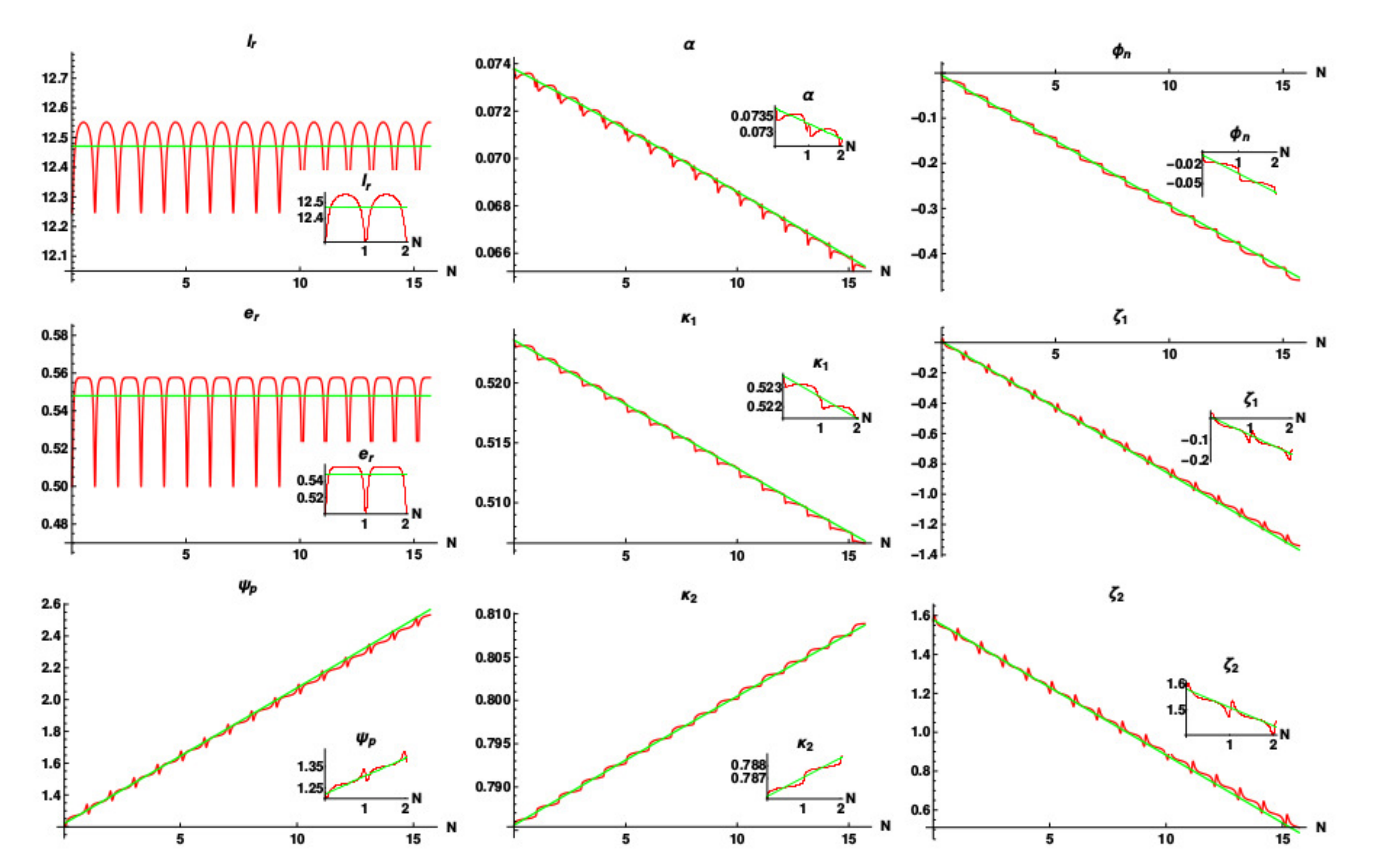} \centering
\caption{The effect of an increased eccentricity $e_{r}\left( 0\right) =0.5$%
, the rest of the parameters being identical to those for Fig. \protect\ref%
{fig01}.}
\label{fig02}
\end{figure*}
\begin{figure*}[th]
\includegraphics[scale=0.53]{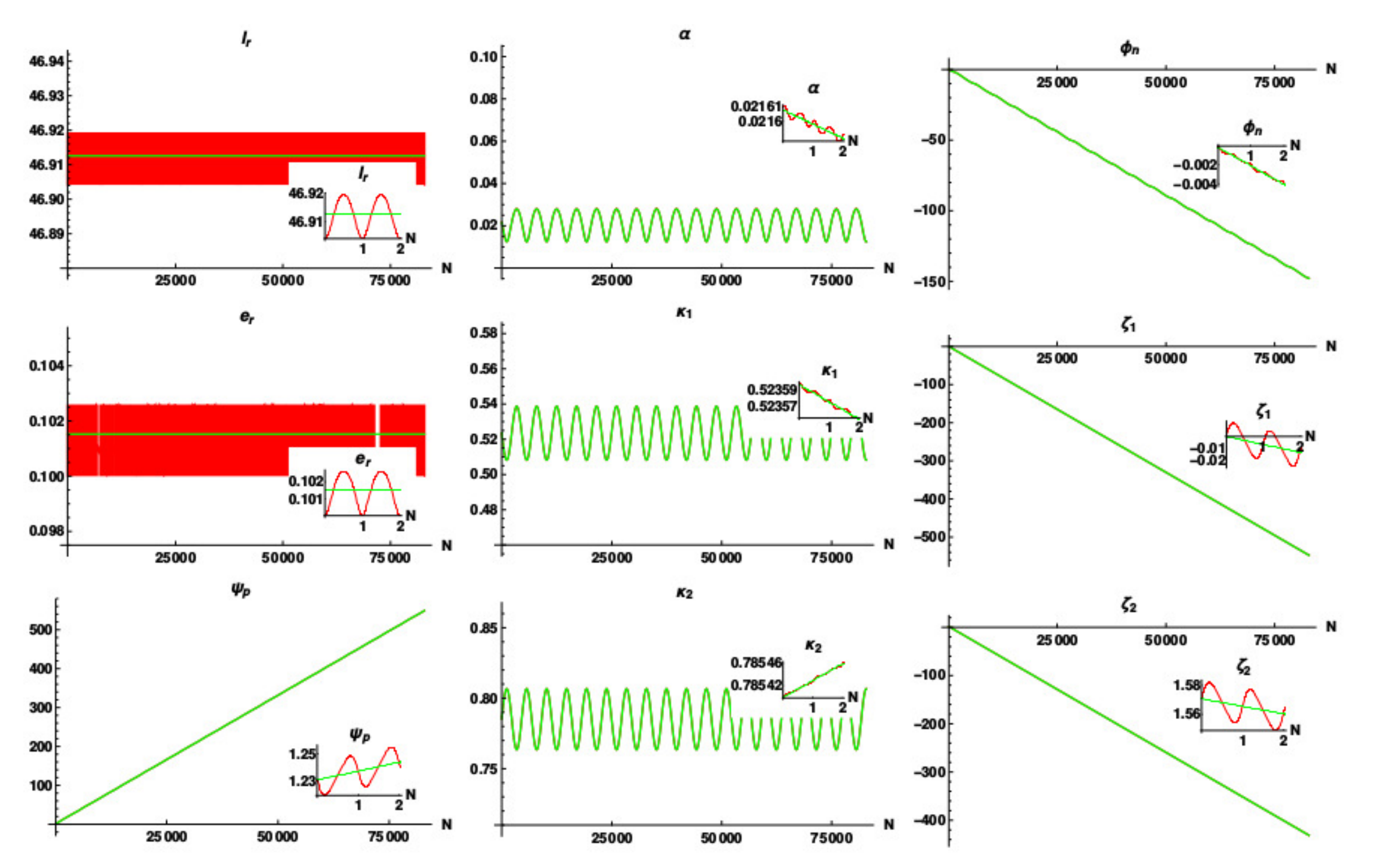} \centering
\caption{The evolutions at a much earlier stage of the inspiral,
characterized by post-Newtonian parameter $\protect\varepsilon \left(
0\right) =0.0005$, the rest of the parameters being identical to those for
Fig. \protect\ref{fig01}. }
\label{fig03}
\end{figure*}
\begin{figure*}[th]
\includegraphics[scale=0.53]{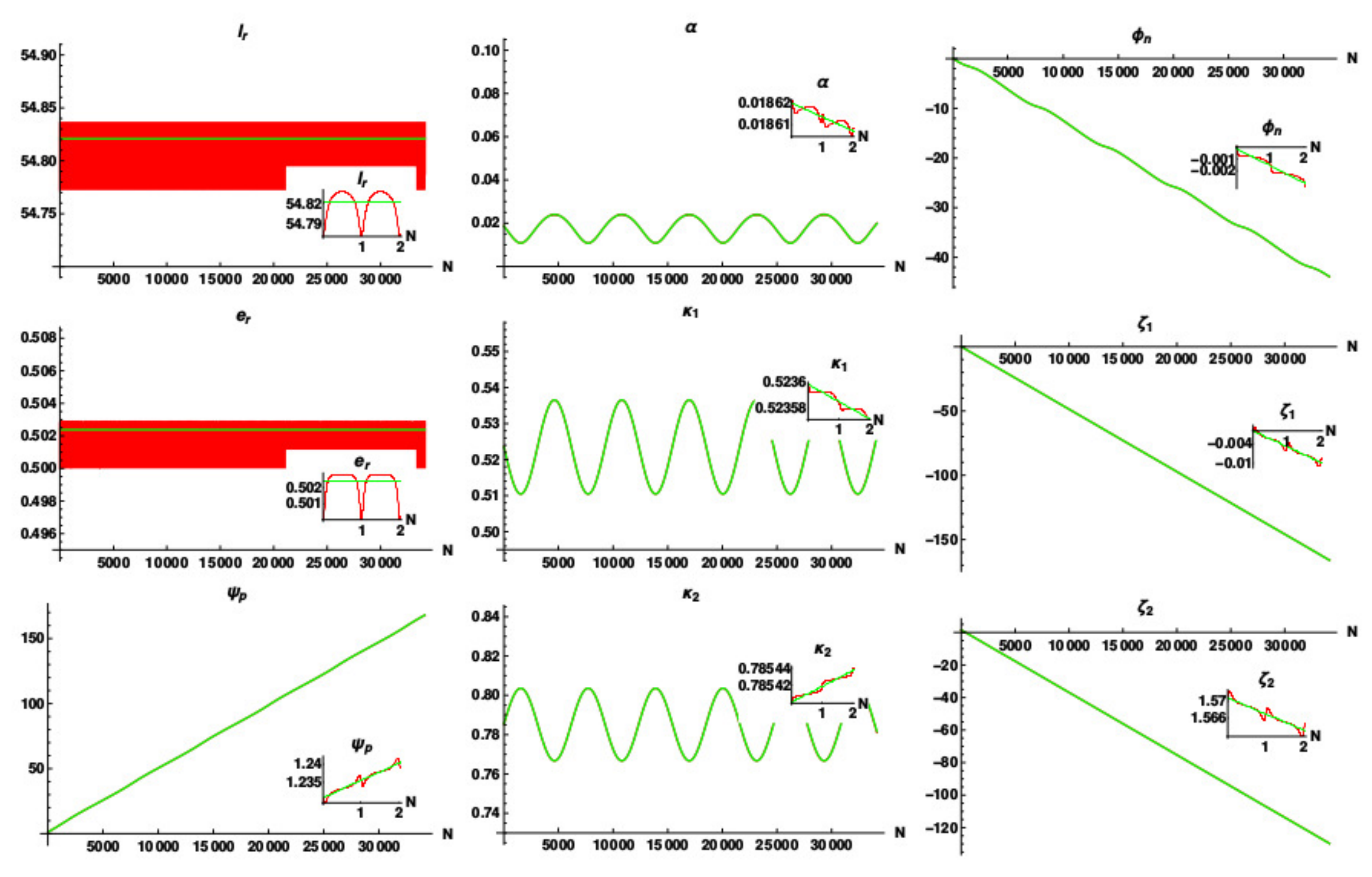} \centering
\caption{The effect of an increased eccentricity $e_{r}\left( 0\right) =0.5$%
, the rest of the parameters being identical to those for Fig. \protect\ref%
{fig03}.}
\label{fig04}
\end{figure*}

The 2PN dynamics of compact binaries (given by Eqs. (36)-(45) of Ref. \cite%
{chameleon}) have four first integrals which are derived in Secs. V.A. and
V.B. of Ref. \cite{chameleon}. The first integrals give the conservation of
total energy and total angular momentum. Two first integrals which express
the direction of total angular momentum and given by Eq. (53) and the ratio
of Eqs. (54) and (55) of Ref. \cite{chameleon} represent two constraints
among the variables occurring in the secular evolution equations. As a
result of this, the secular dynamics given above is subject to two
constraints among the shape variables, the Euler angles, and spin angles,
which will be shown below.

The constraints determining the direction of the total angular momentum to
leading order (see Eqs. (58) and (62) of Ref. \cite{chameleon}) read as%
\begin{equation}
\mathfrak{\bar{l}}_{r}=\sum_{i=1}^{2}\nu ^{2i-3}\chi _{i}\left[ \sin \kappa
_{i}\sin \left( \zeta _{i}+\psi _{p}\right) \cot \alpha -\cos \kappa _{i}%
\right] \ ,  \label{constr1}
\end{equation}%
and%
\begin{equation}
\sum_{i=1}^{2}\nu ^{2i-3}\chi _{i}\sin \kappa _{i}\cos \left( \zeta
_{i}+\psi _{p}\right) =0\ .  \label{constr2}
\end{equation}%
Here $\mathfrak{l}_{r}\left( \chi _{p}\right) $ was changed to $\mathfrak{%
\bar{l}}_{r}$, an approximation valid to leading order. These expressions do
not depend explicitly on the variable $\chi _{p}$ characterizing the
position of the body with reduced mass. Only the correction terms $\mathcal{O%
}\left( \mathfrak{\bar{l}}_{r}^{-2}\right) $ to the above equations exhibit
explicit $\chi _{p}$ dependence. Therefore, we expect the constraints (\ref%
{constr1}) and (\ref{constr2}) to be conserved under the secular evolution.

Since $\mathfrak{\bar{l}}_{r}$ is a constant, the time derivative of the
left-hand side of (\ref{constr1}) is zero. A long but straightforward
computation with the application of the secular evolution equations shows
that the time derivative of the right-hand side also vanishes. Similarly we
checked that the time derivative of Eq. (\ref{constr2}) vanishes to 2PN\
order. These prove that Eqs. (\ref{constr1}) and (\ref{constr2}) are first
integrals of the secular evolution equations given in Secs. \ref{shape}--\ref%
{spin}. They can be used to reduce the order of the differential equation
system or for checking the accuracy of the numerical integration. In
addition, Eqs. (\ref{constr1}) and (\ref{constr2}) have to be fulfilled when
setting up initial conditions.

\section{Secular and instantaneous dynamics of black hole binaries compared 
\label{limits}}

\subsection{Conservative timescale}

Because of gravitational radiation, the PN parameter increases during the
inspiral. During the time 
\begin{equation}
\Delta \tau =\frac{5}{256\eta }\left( \varepsilon _{\left( in\right)
}^{-4}-\varepsilon _{\left( out\right) }^{-4}\right) \frac{Gm}{c^{3}}~
\label{Dtau}
\end{equation}%
the PN parameter evolves from $\varepsilon _{\left( in\right) }$ to $%
\varepsilon _{\left( out\right) }$ \cite{MVG}. For numerical evolution
purposes, we define the dimensionless conservative timescale $\mathfrak{%
\tilde{T}}_{cons}$ as a 1\% increase in the PN parameter: 
\begin{equation}
\mathfrak{\tilde{T}}_{cons}=\frac{5}{256\eta }\varepsilon _{\left( in\right)
}^{-4}\left( 1-\frac{1}{1.01^{4}}\right) \approx \frac{0.195}{256\eta }%
\varepsilon _{\left( in\right) }^{-4}~.  \label{taucons}
\end{equation}%
Dividing this by (the leading-order dimensionless) orbital period $\mathfrak{%
\tilde{T}}_{orb}=2\pi \varepsilon ^{-3/2}$ gives the number 
\begin{equation}
N_{\max }=\frac{0.195}{512\pi \eta }\varepsilon _{\left( in\right) }^{-5/2}~
\label{Nmax}
\end{equation}%
of radial periods of the conservative timescale. Evolving numerically for $%
N_{\max }$ periods keeps the error of disregarding the dissipation below $%
1\% $.

\subsection{Accuracy of the secular dynamics}

We check the long-term accuracy of the secular dynamics by a numerical
comparison over $N_{\max }$ periods with the instantaneous evolutions given
by Eqs. (36)-(42) of Ref. \cite{chameleon}. The results are shown on Figs. %
\ref{fig01}--\ref{fig04}. All figures are for $m=20M_{\odot }$, $\nu =1/2$,
and high spin parameter values $\chi _{1}=\chi _{2}=0.9982$. The initial
value for the PN parameter is $\varepsilon \left( 0\right) =10^{-2}$ in
Figs. \ref{fig01} and \ref{fig02}, while $\varepsilon \left( 0\right)
=5\times 10^{-4}$ in Figs. \ref{fig03} and \ref{fig04}. The initial value
for the eccentricity is $e_{r}\left( 0\right) =0.1$ in Figs. \ref{fig01} and %
\ref{fig03} and $e_{r}\left( 0\right) =0.5$ in Figs. \ref{fig02} and \ref%
{fig04}.

For the instantaneous evolutions, the shape parameter $\mathfrak{l}%
_{r}\left( 0\right) $ emerged from the PN parameter $\varepsilon \left(
0\right) $ and eccentricity $e_{r}\left( 0\right) $ cf. Eq. (8) of Ref. \cite%
{chameleon}. Then $\psi _{p}\left( 0\right) $ and $\alpha \left( 0\right) $
were computed from the constraints given by Eq. (53) and the ratio of Eqs.
(54) and (55) of Ref. \cite{chameleon}.

The initial values $\bar{f}\left( 0\right) $ for the secular dynamics were
extracted from the instantaneous evolution during the first orbit in the
following manner:%
\begin{equation}
\bar{f}\left( 0\right) =\bar{f}-\frac{1}{2}\left[ f\left( \mathfrak{T}%
\right) -f\left( 0\right) \right] ~,  \label{secinit}
\end{equation}%
where $\bar{f}$ represents the first orbital average. This method corrects
for the periodic component of the instantaneous motion leading to the proper
initial condition for the secular evolution representing the orbital average.

In the secular dynamics the shape variables $\bar{e}_{r}$ and $\mathfrak{%
\bar{l}}_{r}$ are conserved. Finally, the initial values of $\psi _{p}$ and $%
\alpha $ were computed from the constraints (\ref{constr1}) and (\ref%
{constr2}).

The smaller pictures zoom into the first two radial periods of the
evolution. On the conservative timescale the secular evolution follows
closely the instantaneous evolutions. Certain evolutions even overlap making
them hard to distinguish.

\subsection{Effects of the eccentricity and PN\ parameter}

The comparison of Figs. (\ref{fig01}) and (\ref{fig02}), and also (\ref%
{fig03}) and (\ref{fig04}), yields to the remark that $N_{\max }$ decreases
with increasing eccentricity, whereas comparing the evolutions of distant
binaries in Figs. (\ref{fig02}) and (\ref{fig04}) to the close binary
evolutions (\ref{fig01}) and (\ref{fig03}) yields that $N_{\max }$ decreases
with increasing PN parameter.

While Figs. (\ref{fig03}) and (\ref{fig04}) represent the early stages of
the inspiral, Figs. (\ref{fig01}) and (\ref{fig02}) refer to the very end of
it. While the secular evolution of the spin azimuthal angles $\zeta _{1,2}$
indicate more than $60$ precessional cycles at low eccentricity and more
than $20$ for high eccentricity in the early inspiral regime, at the end of
it, $N_{\max }$ will not reach even one single precessional cycle.
Similarly, the number of nutations (represented by the number of cycles of
the polar angles $\kappa _{i}$ between their maximal and a minimal values)
is $18$ at low eccentricity and $5$ at high eccentricity in the early
inspiral, while at the end of it, $N_{\max }$ covers only a fraction of the
nutational period. The polar angle $\alpha $ and azimuthal angle $\phi _{n}$
of the orbital angular momentum exhibit similar behavior to that of the
spins.

\section{Closed system for the secular spin angle evolutions\label{closed}}

Remarkably, the secular evolution of the spin angles $\kappa _{1}$ and $%
\kappa _{2}$ and $\Delta \zeta \equiv \zeta _{1}-\zeta _{2}$ discussed in
Sec. \ref{secdyn} form a closed subset:%
\begin{eqnarray}
\frac{1}{R}\overline{\frac{d\kappa _{1}}{d\mathfrak{t}}} &=&\left( 1+\nu
-x_{1}\cos \kappa _{1}-\nu w_{2}x_{2}\cos \kappa _{2}\right)  \notag \\
&&\times x_{2}\sin \kappa _{2}\sin \Delta \zeta ~,  \label{eq1}
\end{eqnarray}%
\begin{eqnarray}
\frac{1}{R}\overline{\frac{d\kappa _{2}}{d\mathfrak{t}}} &=&-\left( 1+\nu
^{-1}-x_{2}\cos \kappa _{2}-\nu ^{-1}w_{1}x_{1}\cos \kappa _{1}\right) 
\notag \\
&&\times x_{1}\sin \kappa _{1}\sin \Delta \zeta ~,  \label{eq2}
\end{eqnarray}%
\begin{eqnarray}
\frac{1}{R}\overline{\frac{d\Delta \zeta }{d\mathfrak{t}}} &=&\nu -\nu
^{-1}+\left( 1+2\nu ^{-1}-w_{1}\right.  \notag \\
&&\left. -w_{1}\nu ^{-1}x_{1}\cos \kappa _{1}\right) x_{1}\cos \kappa _{1} 
\notag \\
&&-\left( 1+2\nu -w_{2}-w_{2}\nu x_{2}\cos \kappa _{2}\right) x_{2}\cos
\kappa _{2}  \notag \\
&&-\left( 1+\nu ^{-1}-w_{1}\nu ^{-1}x_{1}\cos \kappa _{1}\right)  \notag \\
&&\times x_{1}\cot \kappa _{2}\sin \kappa _{1}\cos \Delta \zeta  \notag \\
&&+\left( 1+\nu -w_{2}\nu x_{2}\cos \kappa _{2}\right)  \notag \\
&&\times x_{2}\cot \kappa _{1}\sin \kappa _{2}\cos \Delta \zeta  \notag \\
&&-x_{1}x_{2}\left( \frac{\sin \kappa _{2}}{\sin \kappa _{1}}-\frac{\sin
\kappa _{1}}{\sin \kappa _{2}}\right) \cos \Delta \zeta ~,  \label{eq3}
\end{eqnarray}%
where we have introduced the notations%
\begin{equation}
R=\frac{3\eta \pi }{\mathfrak{T}~\mathfrak{\bar{l}}_{r}^{2}}~,~x_{i}=\frac{%
\chi _{i}}{\mathfrak{\bar{l}}_{r}}~.
\end{equation}%
In Appendix \ref{reg}, we prove that the evolutions remain regular across
manifest coordinate singularities.

In the next two sections, we will discuss applications for this closed
system.

\section{Spin flip-flops when one spin dominates over the other\label%
{particular}}

In this section, we consider the case of one spin dominating over the other,
thus $\chi _{2}\ll \chi _{1}$. Under these conditions we will recover
configurations with large spin flip-flop already discussed in the literature
and an additional flip-flop induced by a particular value of the quadrupole
parameter, relevant for neutron star binaries.

Since%
\begin{equation}
\mathcal{O}\left( \overline{\frac{d\kappa _{1}}{d\mathfrak{t}}}\right) /%
\mathcal{O}\left( \overline{\frac{d\kappa _{2}}{d\mathfrak{t}}}\right)
\approx \frac{\nu \chi _{2}}{\chi _{1}}\ll 1~,  \label{kappasevoratio}
\end{equation}%
in this case, $\kappa _{1}$ behaves as a quasiconstant. The system governing 
$\Delta \zeta $ and $\kappa _{2}$ simplifies to 
\begin{equation}
\overline{\frac{d\Delta \zeta }{d\mathfrak{t}}}=A_{S_{1}}+B_{S_{1}}\cot
\kappa _{2}\cos \Delta \zeta ~,  \label{EqDzeta}
\end{equation}%
\begin{equation}
\overline{\frac{d\kappa _{2}}{d\mathfrak{t}}}=B_{S_{1}}\sin \Delta \zeta ~,
\label{Eqkappa2}
\end{equation}%
with coefficients%
\begin{eqnarray}
\frac{A_{S_{1}}}{R} &=&\nu -\frac{1}{\nu }+\left( 1+\frac{2}{\nu }%
-w_{1}\right.  \notag \\
&&\left. -\frac{w_{1}x_{1}}{\nu }\cos \kappa _{1}\right) x_{1}\cos \kappa
_{1}~,  \label{AS1}
\end{eqnarray}%
and%
\begin{equation}
\frac{B_{S_{1}}}{R}=-\left( 1+\frac{1}{\nu }-\frac{w_{1}x_{1}}{\nu }\cos
\kappa _{1}\right) x_{1}\sin \kappa _{1}~.  \label{BS1}
\end{equation}

For the values $\kappa _{1}=\left\{ 0,\pi \right\} $, the angle $\kappa _{2}$
becomes a constant, and $\Delta \zeta =A_{S_{1}}\mathfrak{t}+$constant.

For generic $\kappa _{1}$, the system has fix points given by $\overline{%
d\Delta \zeta }/d\mathfrak{t}=0$ and $\overline{d\kappa _{2}}/d\mathfrak{t}%
=0 $, resulting in either 
\begin{equation}
\Delta \zeta =0~,~\kappa _{2}=\arctan \left( -\frac{B_{S_{1}}}{A_{S_{1}}}%
\right) ~  \label{fixp}
\end{equation}%
or 
\begin{equation}
\Delta \zeta =\pi ~,~\kappa _{2}=\arctan \left( \frac{B_{S_{1}}}{A_{S_{1}}}%
\right) ~.  \label{fixp2}
\end{equation}

In the rest of the cases [other than $\kappa _{1}=\left\{ 0,\pi \right\} $
or the fix points (\ref{fixp}) or (\ref{fixp2})], we derive the following
second-order differential equation from Eqs. (\ref{EqDzeta})--(\ref{Eqkappa2}%
), 
\begin{equation}
\overline{\frac{d^{2}}{d\mathfrak{t}^{2}}}\left( \sin \kappa _{2}\sin \Delta
\zeta \right) +\Omega _{S_{1}}^{2}\sin \kappa _{2}\sin \Delta \zeta =0~,
\label{ddsinkappasinDzeta}
\end{equation}%
where%
\begin{equation}
\Omega _{S_{1}}=\sqrt{A_{S_{1}}^{2}+B_{S_{1}}^{2}}~.
\end{equation}%
Equation (\ref{ddsinkappasinDzeta}) represents a harmonic oscillator, and
its solution reads%
\begin{equation}
\sin \kappa _{2}\sin \Delta \zeta =K_{1}\cos \left( \Omega _{S_{1}}\mathfrak{%
\bar{t}}+D_{S_{1}}\right) ~,  \label{sol1S}
\end{equation}%
with integration constants $\left\vert K_{1}\right\vert \leq 1$ and $%
D_{S_{1}}$. Then, from (\ref{Eqkappa2}), we obtain%
\begin{equation}
\cos \kappa _{2}=-\frac{K_{1}B_{S_{1}}}{\Omega _{S_{1}}}\sin \left( \Omega
_{S_{1}}\mathfrak{\bar{t}}+D_{S_{1}}\right) +K_{2}~,  \label{sol2S}
\end{equation}%
where $\left\vert K_{2}\right\vert \leq 1$ is an additional integration
constant. Since the system (\ref{EqDzeta})--(\ref{Eqkappa2}) of two
first-order differential equations admits only two integration constants, $%
K_{1}$, $K_{2}$, and $D_{S_{1}}$ are not independent. Indeed, by taking the
derivative of Eq. (\ref{sol1S}) and using Eqs. (\ref{EqDzeta})--(\ref%
{Eqkappa2}) and (\ref{sol2S}), we find that%
\begin{equation}
A_{S_{1}}\sin \kappa _{2}\cos \Delta \zeta =\frac{A_{S_{1}}^{2}}{B_{S_{1}}}%
\cos \kappa _{2}-\frac{\Omega _{S_{1}}^{2}}{B_{S_{1}}}K_{2}~.  \label{relint}
\end{equation}%
Hence $A_{S_{1}}=0$ also implies $K_{2}=0$. We introduce a new constant $C$
as%
\begin{equation}
K_{2}=A_{S_{1}}C~
\end{equation}%
with arbitrary value for $A_{S_{1}}=0$ and from (\ref{relint}) and (\ref%
{sol1S})--(\ref{sol2S}) 
\begin{equation}
K_{1}^{2}=1-\Omega _{S_{1}}^{2}C^{2}
\end{equation}%
otherwise. Hence the solutions (\ref{sol1S}) and (\ref{sol2S}) can be
rewritten as 
\begin{equation}
\cos \kappa _{2}=A_{S_{1}}C+\frac{\epsilon B_{S_{1}}}{\Omega _{S_{1}}}\sqrt{%
1-\Omega _{S_{1}}^{2}C^{2}}\sin \left( \Omega _{S_{1}}\mathfrak{\bar{t}}%
+D_{S_{1}}\right) ~,  \label{1}
\end{equation}%
\begin{equation}
\sin \kappa _{2}\sin \Delta \zeta =-\epsilon \sqrt{1-\Omega _{S_{1}}^{2}C^{2}%
}\cos \left( \Omega _{S_{1}}\mathfrak{\bar{t}}+D_{S_{1}}\right) ~,  \label{2}
\end{equation}%
where $\epsilon =\pm 1$ and $-\epsilon $ gives the sign of $K_{1}$.\footnote{%
Note that the reverse case $\chi _{1}\ll \chi _{2}$ can be obtained by the
following notational changes: $\chi _{1}\rightarrow \chi _{2}$, $\nu
\rightarrow \nu ^{-1}$, $w_{1}\rightarrow w_{2}$, $\kappa _{1}\rightarrow
\kappa _{2}$, and $\Delta \zeta \rightarrow -\Delta \zeta $.}

\begin{figure*}[th]
\includegraphics[scale=0.45]{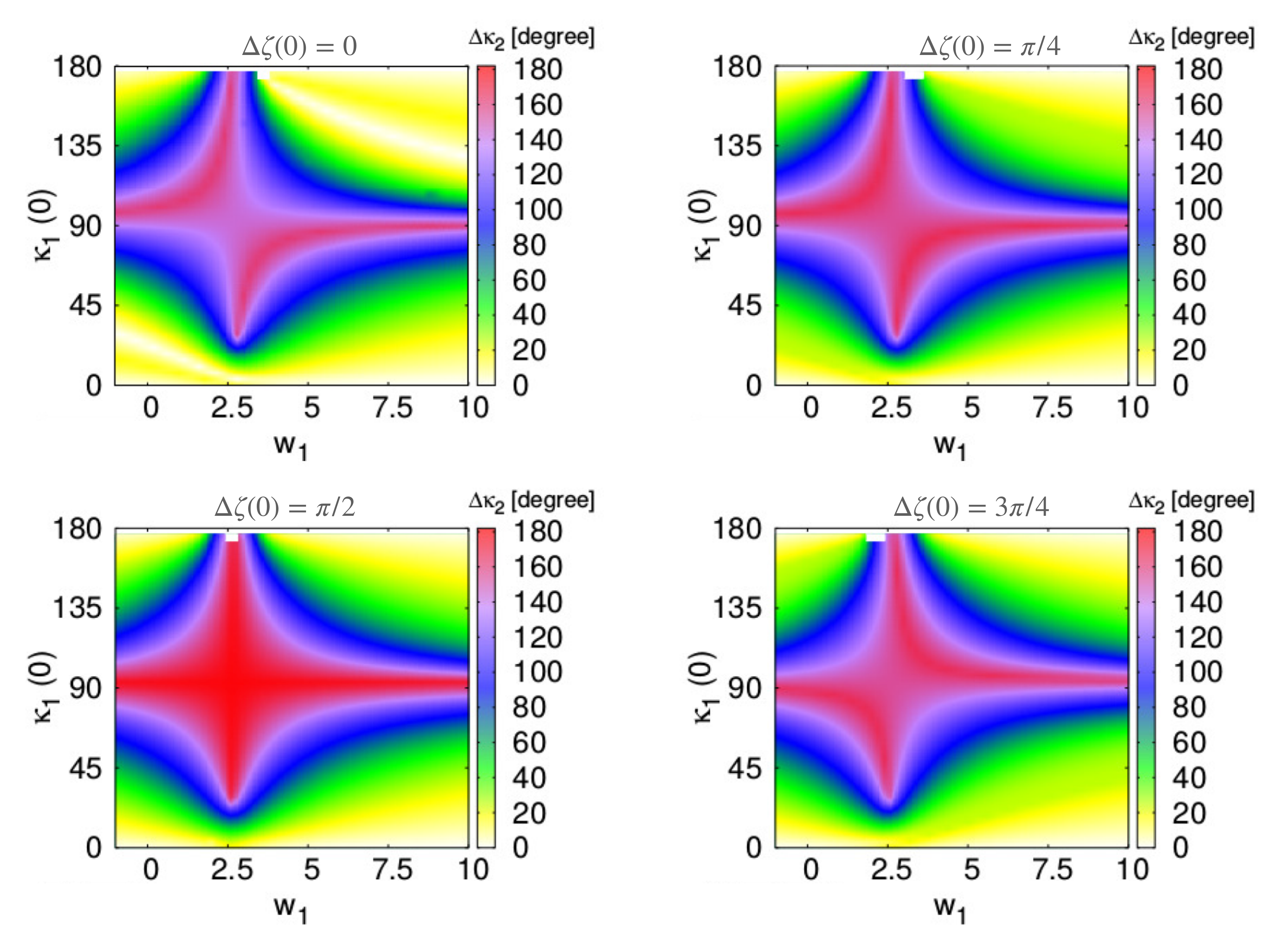}
\caption{The color map represents the flip-flop angle $\Delta \protect\kappa %
_{2}$ as a function of the initial polar angle $\protect\kappa _{1}\left(
0\right) $ and the quadrupole parameter $w_{1}$ of the dominant spin ($%
\protect\chi _{1}\gg \protect\chi _{2}$). From the top left to lower right
panel, $\Delta \protect\zeta \left( 0\right) $ is ${0}$, ${\protect\pi /4}$, 
${\protect\pi /2}$, and ${3\protect\pi /4}$, respectively. The additional
parameters of the binary are $m_{1}=m_{2}=1.4M_{\odot }$, $\bar{e}_{r}=0.1$, 
$\bar{\protect\varepsilon}=0.0001$, $\protect\chi _{1}=0.95$, $\protect\chi %
_{2}=0.05$, and $\protect\kappa _{2}\left( 0\right) =\protect\pi /10$. In
the small white rectangle regions, $\Delta \protect\kappa _{2}$ does not
reach its maximum within the conservative timescale. The flip-flop is large
in the red diamond-shaped regions. }
\label{fig05}
\end{figure*}

From Eq. (\ref{1}), the maximal variation of $\cos \kappa _{2}$ during the
evolution is%
\begin{eqnarray}
\left\vert \Delta \cos \kappa _{2}\right\vert &=&\left\vert \frac{2B_{S_{1}}%
}{\Omega _{S_{1}}}\sqrt{1-\Omega _{S_{1}}^{2}C^{2}}\right\vert  \notag \\
&=&\left\vert 2\sqrt{\frac{B_{S_{1}}^{2}}{A_{S_{1}}^{2}+B_{S_{1}}^{2}}%
-B_{S_{1}}^{2}C^{2}}\right\vert ~.  \label{Dcoskappa2}
\end{eqnarray}%
This is the largest when%
\begin{equation}
A_{S_{1}}^{2}<<B_{S_{1}}^{2}~.  \label{cond}
\end{equation}%
As $x_{1}=\chi _{1}/\mathfrak{\bar{l}}_{r}$ and $1/\mathfrak{\bar{l}}_{r}$
represents 1/2PN order (Ref. \cite{chameleon}), a necessary condition for
Eq. (\ref{cond}) is $\nu \approx 1$, since in this case the leading-order
term of $A_{S_{1}}^{2}$ is negligible. Further the condition (\ref{cond}),
implying a large change in the polar angle of the smaller spin cf. Eq. (\ref%
{Dcoskappa2}), holds in two cases,%
\begin{equation}
i)~\cos \kappa _{1}=\mathcal{O}\left( \frac{1}{\mathfrak{\bar{l}}_{r}}%
\right) ~,  \label{BH}
\end{equation}%
and%
\begin{equation}
ii)~1+\frac{2}{\nu }-w_{1}-\frac{w_{1}x_{1}}{\nu }\cos \kappa _{1}=\mathcal{O%
}\left( \frac{1}{\mathfrak{\bar{l}}_{r}}\right) ~.  \label{NS}
\end{equation}%
The condition (\ref{BH}) means that the larger spin vector is almost
perpendicular to $\mathbf{L}_{N}$ (with a deviation of 1/4PN\ order
allowed). This configuration, which results in large flip-flops of the
smaller spin was analyzed first in Ref. \cite{flipflop} for quasicircular
orbits. By contrast, as $\kappa _{1}$\ is contained in a $\mathcal{O}\left(
1/\mathfrak{\bar{l}}_{r}\right) $ term of Eq. (\ref{NS}), the second
condition could hold for a wide range of angles $\kappa _{1}$ (i.e. almost
independently of the direction of the dominant spin), provided the binary
component with dominant spin has a quadrupole moment $w_{1}\approx 3$. This
situation can be relevant for neutron star - neutron star binary systems
where one of the binary components is spinning much faster than the other.
Both flip-flop situations arising under i) and ii) are represented on Fig. %
\ref{fig05} in a\ combined fashion as the red diamond-shaped regions. Case
i) occurs along the horizontal axis, while case ii) along the vertical one. 

\section{Binaries with a black hole and a gravastar, another black hole or
boson star companion\label{boson}}

\subsection{Black hole - boson star binaries}

In this subsection we discuss analytically the case of black hole - boson
star binaries with $w_{1}=1$ and $w_{2}\approx 100$, also equal masses and
at large separations, such that $x_{1}\approx x_{2}\ll 1$ (as they are of
the order of 1/2PN order; also we assume $w_{2}x_{2}\ll 1$). The closed
system (\ref{eq1})-(\ref{eq3}) can be approximated as%
\begin{equation}
\frac{1}{R}\overline{\frac{d\kappa _{1}}{d\mathfrak{t}}}=2x_{2}\sin \kappa
_{2}\sin \Delta \zeta ~,  \label{eq1b}
\end{equation}%
\begin{equation}
\frac{1}{R}\overline{\frac{d\kappa _{2}}{d\mathfrak{t}}}=-2x_{1}\sin \kappa
_{1}\sin \Delta \zeta ~,  \label{eq2b}
\end{equation}%
\begin{eqnarray}
\frac{1}{R}\overline{\frac{d\Delta \zeta }{d\mathfrak{t}}} &=&2x_{1}\cos
\kappa _{1}+w_{2}x_{2}\cos \kappa _{2}  \notag \\
&&-2x_{1}\cot \kappa _{2}\sin \kappa _{1}\cos \Delta \zeta  \notag \\
&&+2x_{2}\cot \kappa _{1}\sin \kappa _{2}\cos \Delta \zeta ~.  \label{eq3b}
\end{eqnarray}

The first two equations give 
\begin{equation}
\overline{\frac{d\cos \kappa _{1}}{d\cos \kappa _{2}}}=-\frac{x_{2}}{x_{1}}~,
\end{equation}%
or 
\begin{equation}
\cos \kappa _{2}=\frac{x_{1}}{x_{2}}\left( A-\cos \kappa _{1}\right) ~,
\label{coskaka}
\end{equation}%
with $A$ a suitable constant to render $\cos \kappa _{2}$\ in the allowed
range. If $\cos \kappa _{2}\neq 0$, the first term of Eq. (\ref{eq3b}) can
also be dropped.

In the particular case $A=0~$and $x_{2}=x_{1}$ Eq. (\ref{coskaka}) yields $%
\kappa _{2}=\pi -\kappa _{1}$.

This is a highly symmetrical configuration, with the evolutions of the two
spin polar angles occurring symmetrically to the orbital plane. In this
setup, Eqs. (\ref{eq1b})--(\ref{eq3b}) reduce to%
\begin{equation}
\frac{1}{R}\overline{\frac{d\kappa _{1}}{d\mathfrak{t}}}=2x_{1}\sin \kappa
_{1}\sin \Delta \zeta ~,  \label{eq1c}
\end{equation}%
\begin{equation}
\frac{1}{R}\overline{\frac{d\Delta \zeta }{d\mathfrak{t}}}=-w_{2}x_{1}\cos
\kappa _{1}~.  \label{eq3c}
\end{equation}%
These suitably combined integrate into%
\begin{equation}
\cos \Delta \zeta =\ln \left\vert C\sin ^{\frac{w_{2}}{2}}\kappa
_{1}\right\vert ~,  \label{eq3d}
\end{equation}%
with $C$ an integration constant. Inserting this into Eq. (\ref{eq1c}), we
obtain an ordinary differential equation%
\begin{eqnarray}
\frac{1}{R}\overline{\frac{d\kappa _{1}}{d\mathfrak{t}}} &=&\pm 2x_{1}\sin
\kappa _{1}  \notag \\
&&\times \left[ 1-\left( \ln \left\vert C\sin ^{\frac{w_{2}}{2}}\kappa
_{1}\right\vert \right) ^{2}\right] ^{1/2}~,
\end{eqnarray}%
with formal solution%
\begin{eqnarray}
&&2x_{1}R\left( \mathfrak{t-t}_{0}\right)  \notag \\
&=&\pm \int \frac{d\kappa _{1}}{\left[ 1-\left( \ln \left\vert C\sin ^{\frac{%
w_{2}}{2}}\kappa _{1}\right\vert \right) ^{2}\right] ^{1/2}\sin \kappa _{1}}%
~,  \label{eq1d}
\end{eqnarray}%
where $\mathfrak{t}_{0}$ is another constant. Hence, the time evolution of
the spin polar angles is given by Eq. (\ref{eq1d}), while the evolution of
the difference of their azimuthal angles is given by Eq. (\ref{eq3d}).

\subsection{Comparing flip-flops in black hole - black hole, black hole -
gravastar and black hole - boson star binaries of equal mass}

\begin{figure*}[th]
\includegraphics[scale=0.68]{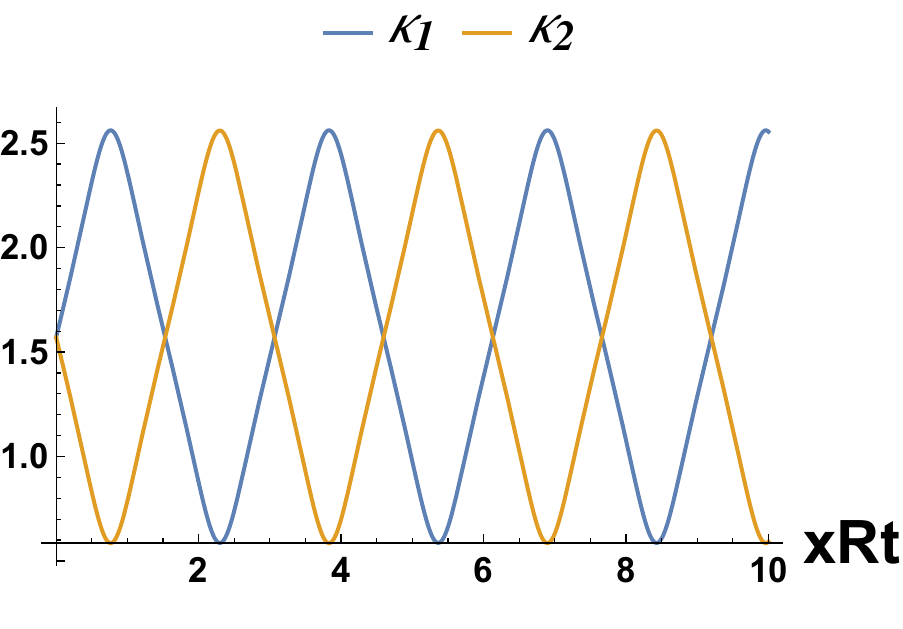} \hspace*{5mm} %
\includegraphics[scale=0.5]{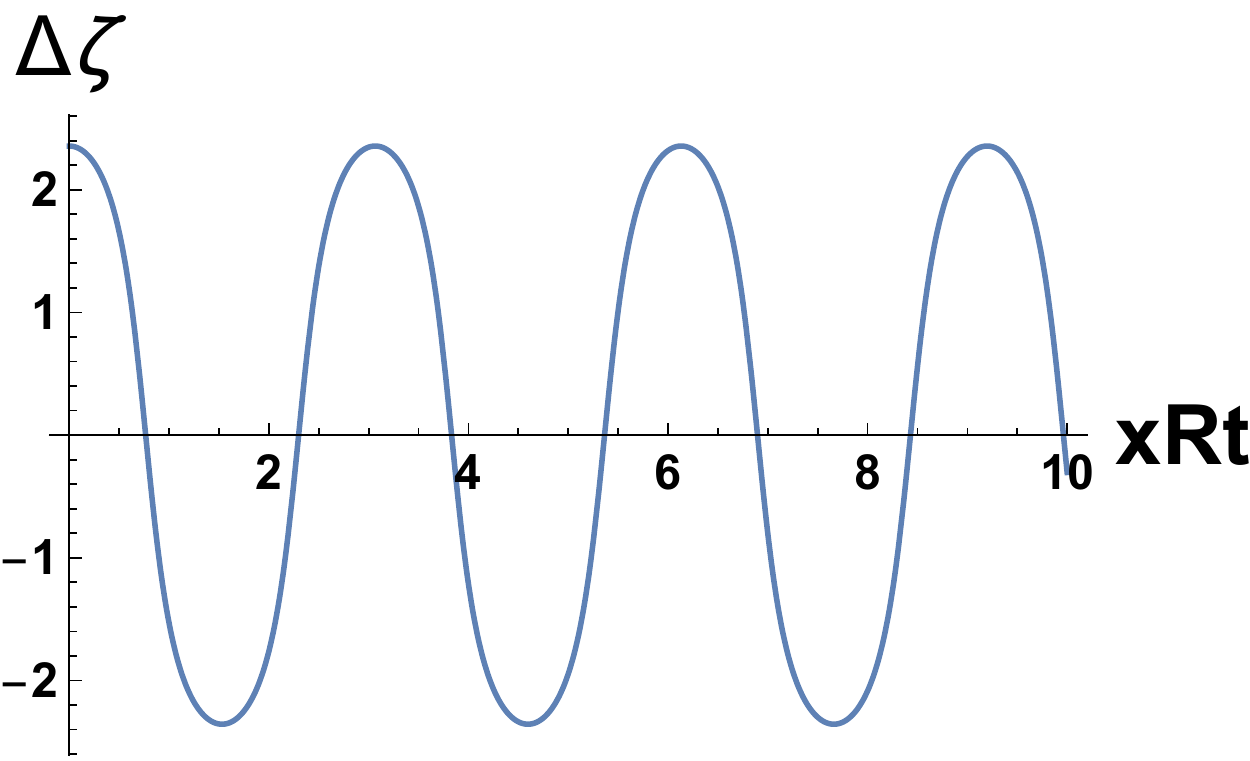} \vspace*{5mm} %
\includegraphics[scale=0.68]{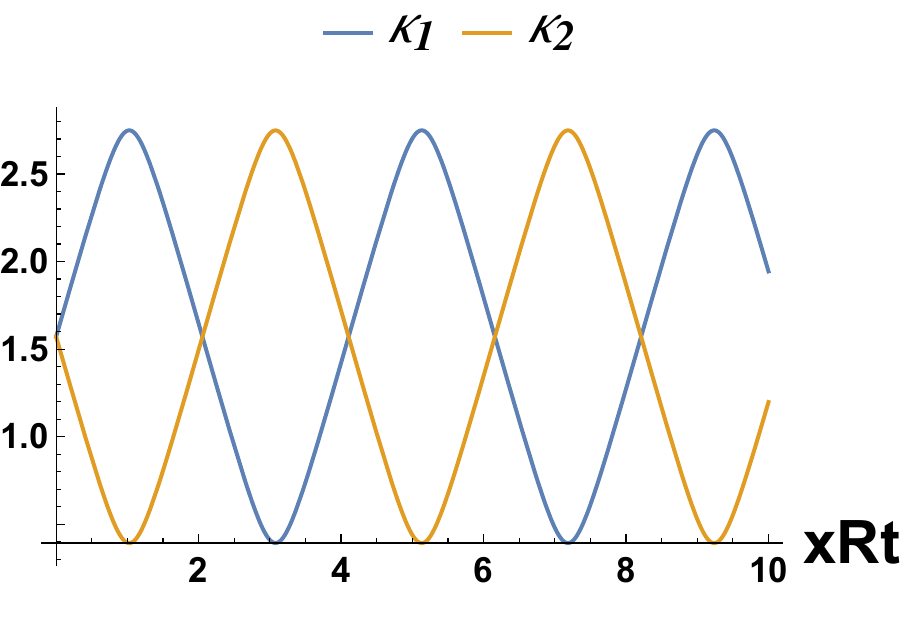} \hspace*{5mm} %
\includegraphics[scale=0.5]{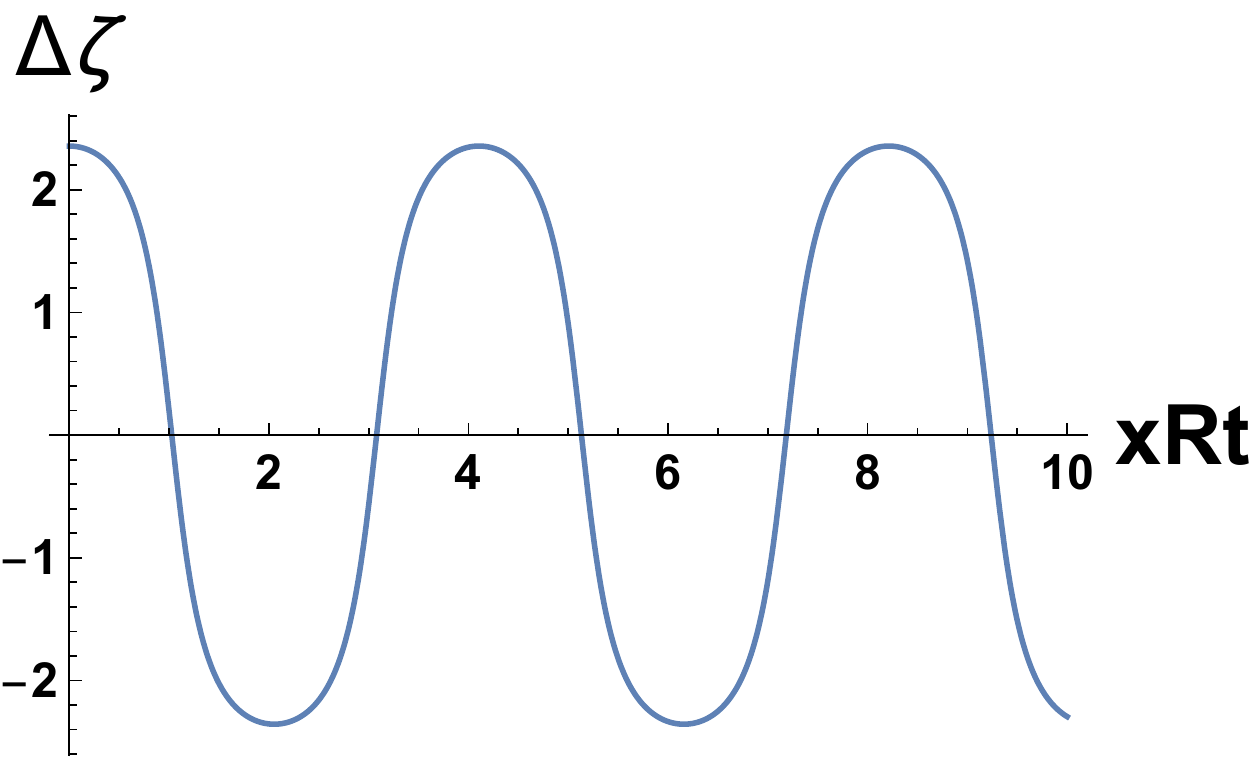} \vspace*{5mm} %
\includegraphics[scale=0.68]{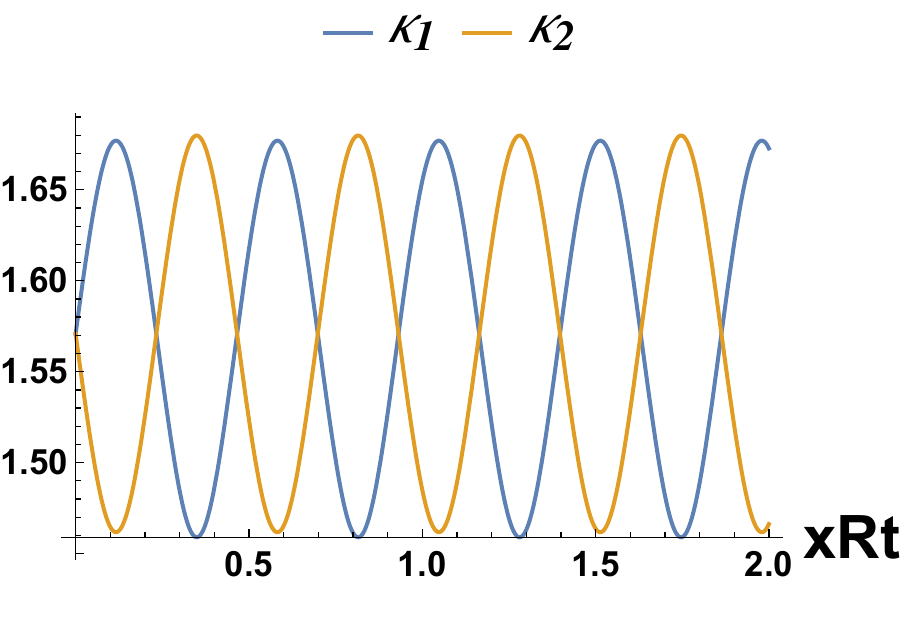} \hspace*{5mm} %
\includegraphics[scale=0.5]{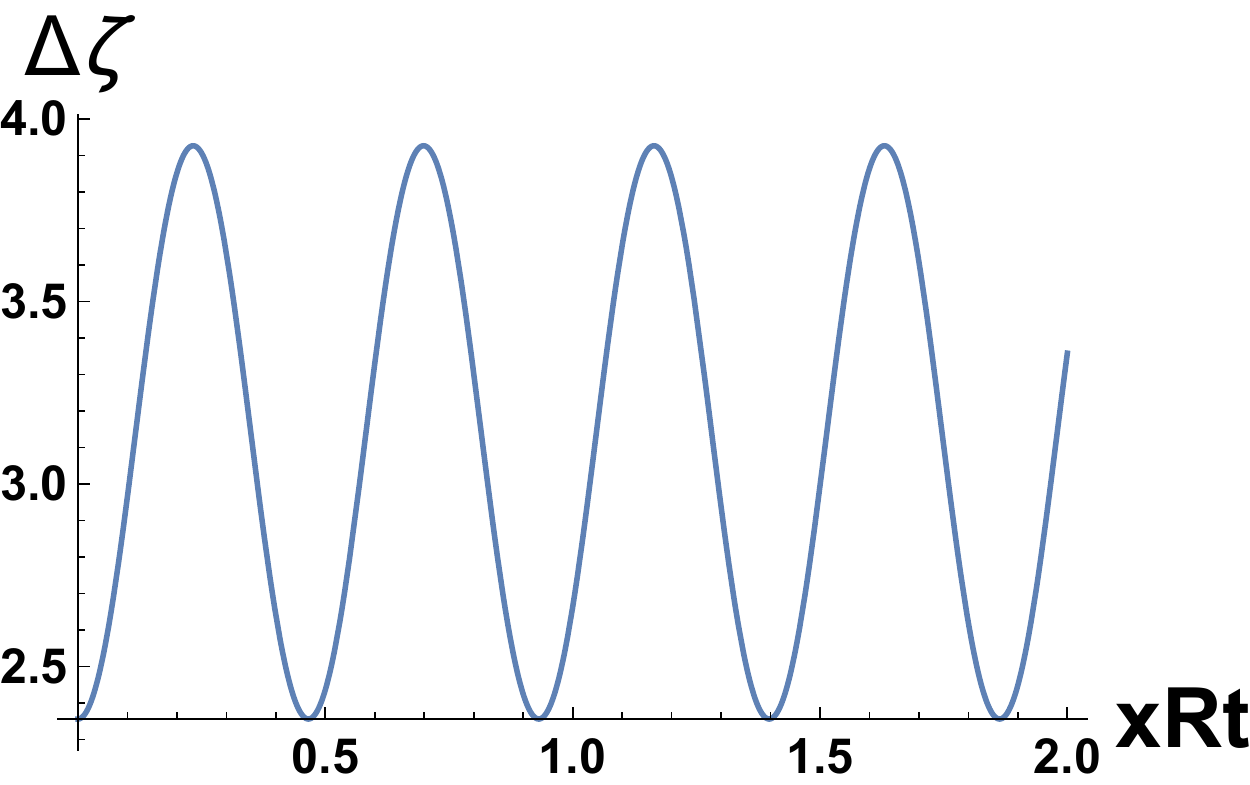}
\caption{The evolution of the spin polar angles (left column) and of the
difference of their azimuthal angle (right column) as function of the
dimensionless time $\mathfrak{t}$ in units of $xR$, for binaries consisting
of a black hole and either a gravastar with $w=-0.8$ (first row), another
black hole (second row) or a boson star with $w=100$ (third row). The mass
ratio is $\protect\nu =1$ in all cases. The plots are for the PN\ parameter $%
\bar{\protect\varepsilon}=0.0001$, eccentricity $\bar{e}_{r}=0.1$, the
parameters $x\equiv x_{1}=x_{2}=0.01$, and initial spin orientations $%
\protect\kappa _{1}\left( 0\right) =\protect\kappa _{2}\left( 0\right) =%
\protect\pi /2$, $\Delta \protect\zeta \left( 0\right) =3\protect\pi /4$.
The polar angles exhibit a \ flip-flopping behavior, with both the frequency
and amplitude depending on the quadrupole parameter of the compact
companion. The difference of the azimuthal angles is also sensitive to $w$.}
\label{fig06}
\end{figure*}

We wish to study here the effect of the quadrupole parameter of the
companion compact object to a black hole in the spin flip-flop. The masses
of all compact objects were chosen equal; hence, $\nu =1$. We monitored the
evolution at PN\ parameter $\bar{\varepsilon}=0.0001$ and eccentricity $\bar{%
e}_{r}=0.1$, leading to $\mathfrak{\bar{l}}_{r}=99.5$. We chose the spin
magnitudes $\chi _{1}=\chi _{2}=0.95$, generating the parameters $x\equiv
x_{1}=x_{2}=0.01$. For initial values of the spin angles we picked $\kappa
_{1}\left( 0\right) =\kappa _{2}\left( 0\right) =\pi /2$ (hence, the spins
lying in the orbital plane), separated by an azimuthal angle difference of $%
\Delta \zeta \left( 0\right) =3\pi /4$. With these values we monitored the
spin angle evolutions for the three distinct binary systems. The results are
represented in Fig. \ref{fig06}.

In all cases, the spin polar angles exhibit the flip-flopping behavior, its
amplitude and frequency being affected by the nature of the companion
(through the value of the quandrupolar parameter $w$). We found that the
largest amplitude and period occur when the companion is an identical black
hole. Both for gravastar and boson star companions, the amplitude and period
of the flip-flop decrease. The flip-flop frequency is largely increased for
boson star companions. As concerning the difference of the azimuthal angles $%
\Delta \zeta \,$, the evolutions are similar for low values of $w$, thus for
gravastar and black hole companions. The spins of the binary components
exhibit precessions during which they overpass each other periodically ($%
\Delta \zeta $ evolves through a sequence of positive and negative values).
Nevertheless when the companion is a boson star, this swinging behavior
disappears ($\Delta \zeta $ remains positive). The amplitude of the
periodical evolution in $\Delta \zeta $ is also decreased.

\section{Conclusions\label{concludingr}}

In this paper we derived the conservative secular evolution of precessing
compact binaries on eccentric orbits, to second post-Newtonian--order
accuracy, with leading-order spin-orbit, spin-spin and mass
quadrupole-monopole contributions included. Our approach relies on employing
a properly chosen set of dimensionless variables, advanced in Ref. \cite%
{chameleon} and a method of averaging over the radial period. The secular
dynamics emerged by applying this to the instantaneous dynamics discussed in
Ref. \cite{chameleon}. The inclusion of the mass quadrupole parameter allows
to apply the formalism for binaries with arbitrary compact components, like
black holes, neutron stars, boson stars or gravastars.

The derived secular dynamics generalizes previous results from the
literature. In Ref. \cite{GPV3} the dynamics was only expressed to 1.5 PN
order by employing different shape variables. In Ref. \cite{ACST}, the
precession was examined with leading-order SO and SS effects for orbits with
negligible eccentricity. Reference \cite{damourderuelle} investigated the PN
dynamics with eccentricity, but without spins. Reference \cite{WOEdyn}
discussed the dynamics of small mass ratio binaries with only the smaller
body having spin. The seminal review on gravitational radiation from compact
binary sources by Blanchet \cite{blanchet} discusses the SO effect in its
last section, but omits the SS and QM contributions to the dynamics. The
secular dynamics where the leading-order SO, SS, and QM coupling are
included is investigated analitically in Ref. \cite{Racine} only for black
holes.

The secular evolution equations emerged as a closed system of first-order
differential equations, which in contrast with the instantaneous evolutions
presented in Ref. \cite{chameleon}, is autonomous. The dependent variables
are the polar ($\kappa _{1}$ and $\kappa _{2}$) and the azimuthal angles ($%
\zeta _{1}$ and $\zeta _{2}$) of the spin vectors, the angles $\alpha $ and $%
\phi _{n}$ giving the orientation of the orbital angular momentum vector,
together with the periastron angle $\psi _{p}$, the dimensionless magnitude
of the orbital angular momentum $\mathfrak{l}_{r}$ and the eccentricity $%
e_{r}$. Over the conservative timescale the secular dynamics can be regarded
as some sort of smoothed-out intantaneous evolution, as illustrated on Figs. %
\ref{fig01}--\ref{fig04}.

Moreover we have shown that the spin polar angles and the difference of
their azimuthal angles in the system defined by the orbital plane evolve
according to a closed subsystem of the secular dynamics. In spite of the
apparent singularity of spherical polar coordinates, the evolutions remains
well defined through aligned configurations. We studied in detail this
closed subsystem in two significant cases. First we assumed that the masses
are comparable, but one of the spins dominates over the other. In this case
we i) derived analytically that large flip-flops of the smaller spin emerge
when the larger spin is almost coplanar with the orbit (a known result) and
ii) found new flip-flop configurations arising for the quadrupole parameter $%
w_{1}\approx 3$ of the neutron star with dominant spin.

We also studied black hole - boson star binaries. In this case, the huge
quadrupolar parameter of the boson star allows for significant
simplification of the closed subsystem, allowing us to derive a formal
analytical solution.

Finally, we analyzed the evolutions of the spin angles numerically by
comparing the cases when the black hole companion is either a gravastar,
another black hole or a boson star with identical mass. We found that the
amplitude and period of the flip-flop is maximal, when the companion is a
black hole. In the case of a boson star companion the frequency of the
flip-flop increased significantly. The precession of the spins is also
sensitive to the quadrupolar parameter of the companion. While in the case
of gravastars and black holes a swinging-type evolution occurs, when the
spins of the components regularly overpass each other, their sequence is
conserved when the companion is a boson star.

In a related paper \cite{KGstab} we will discuss the equilibrium of the spin
configurations and their linear stability in precessing compact binaries
with black hole, neutron star, gravastar, or boson star components.

\section*{Acknowledgements}

This work was supported by the Hungarian National Research Development and
Innovation Office (NKFIH) in the form of the Grant No. 123996 and has been
carried out in the framework of COST actions CA16104 (GWverse) and CA18108
(QG-MM) supported by COST (European Cooperation in Science and Technology).
Z. K. was further supported by the János Bolyai Research Scholarship of the
Hungarian Academy of Sciences and by the ÚNKP-20-5--New National Excellence
Program of the Ministry for Innovation and Technology through its National
Research, Development and Innovation Fund. In the early stages of this work
L. Á. G. was supported by the European Union and the State of Hungary,
cofinanced by the European Social Fund in the framework of TÁMOP 4.2.4.
A/2-11-/1-2012-0001 \textquotedblleft National Excellence
Program.\textquotedblright 

\appendix

\section{The PN expansion of the radial period\label{appendix1}}

When averaging instantaneous variables over one radial orbit in order to
obtain their secular counterpart, the expression of $\dot{\chi}_{p}$ given
by Eq. (43) of Ref. \cite{chameleon} is needed, as explained in Sec. \ref%
{averaging}, with both its Newtonian and PN contributions expressible by the
true anomaly $\chi _{p}$ and the shape variables $\mathfrak{l}_{r}\left(
\chi _{p}\right) $ and $e_{r}\left( \chi _{p}\right) $ alone. Hence $%
\mathfrak{l}_{r}\left( \chi _{p}\right) $ and $e_{r}\left( \chi _{p}\right) $
will be required to 2PN-order accuracy, while the rest of the orbital
elements $\psi _{p}$, $\alpha $, and $\phi _{n}$, and spin angles $\kappa
_{i}$ and $\zeta _{i}$ ($i=1,2$) only to leading order, where they are
constants.

In this Appendix first we derive the $\chi _{p}$ dependence of the
dimensionless orbital angular momentum $\mathfrak{l}_{r}$ and the
dimensionless orbital eccentricity $e_{r}$ to 2PN order, in terms of their
values at the periastron (characterized by the true anomaly $\chi _{p}=0$).
Next, employ these expressions to compute the radial period to 2PN accuracy.
The derivation of the time-averaged values $\bar{\mathfrak{l}}_{r}$ and $%
\bar{e}_{r}$ over the radial period follows, again to 2PN accuracy, with the
inclusion of all spin and mass quadrupole effects to this order. This
enables us to express the shape variables evaluated at the periastron in
terms of the corresponding averaged quantities. This is employed for
rewriting the radial period as a PN expansion in terms of averaged
quantities. At the end of the Appendix we also give a similar expansion of
the averaged PN parameter.

\subsection{$\protect\chi _{p}$ dependence of $\mathfrak{l}_{r}$\label{lrevo}%
}

A lower index $0$ indicates values taken at $\chi _{p}=0$:%
\begin{eqnarray}
\mathfrak{l}_{r}\left( \chi _{p}=0\right) &=&\mathfrak{l}_{r0}~, \\
e_{r}\left( \chi _{p}=0\right) &=&e_{r0}~.
\end{eqnarray}%
The expressions $\mathfrak{\dot{l}}_{r}$ and $\dot{\chi}_{p}$ given by Eqs.
(36) and (43) of Ref. \cite{chameleon} allow for deriving 
\begin{eqnarray}
\mathfrak{l}_{r}\left( \chi _{p}\right) &=&\mathfrak{l}_{r0}+\int_{0}^{\chi
_{p}}\frac{\mathfrak{\dot{l}}_{r}}{\dot{\chi}_{p}}d\chi _{p}  \notag \\
&=&\mathfrak{l}_{r0}+\frac{\mathfrak{l}_{rPN}\left( \chi _{p}\right) }{%
\mathfrak{l}_{r0}}+\frac{\mathfrak{l}_{rSO}\left( \chi _{p}\right) }{%
\mathfrak{l}_{r0}^{2}}  \notag \\
&&+\frac{\mathfrak{l}_{rQM}\left( \chi _{p}\right) }{\mathfrak{l}_{r0}^{3}}+%
\frac{\mathfrak{l}_{rSS}\left( \chi _{p}\right) }{\mathfrak{l}_{r0}^{3}} 
\notag \\
&&+\frac{\mathfrak{l}_{r2PN}\left( \chi _{p}\right) }{\mathfrak{l}_{r0}^{3}}%
~,
\end{eqnarray}%
with

\begin{equation}
\mathfrak{l}_{rPN}\left( \chi _{p}\right) =2\left( 2-\eta \right)
e_{r0}\left( 1-\cos \chi _{p}\right) ~,
\end{equation}%
\begin{equation}
\mathfrak{l}_{r2PN}\left( \chi _{p}\right)
=\sum_{k=0}^{4}\sum_{l=0}^{2}L_{kl}^{2PN}\sin ^{2l}\chi _{p}\cos ^{k}\chi
_{p}~,  \label{lrchi2PN}
\end{equation}%
\begin{eqnarray}
\mathfrak{l}_{rSO}\left( \chi _{p}\right) &=&-\frac{\eta e_{r0}}{2}\left(
\cos \chi _{p}-1\right)  \notag \\
&&\times \sum_{k=1}^{2}\left( 4^{2k-3}+3\right) \chi _{k}\cos \kappa _{k}~,
\end{eqnarray}%
\begin{eqnarray}
\mathfrak{l}_{rSS}\left( \chi _{p}\right) &=&\eta \chi _{1}\chi _{2}\sin
\kappa _{1}\sin \kappa _{2}  \notag \\
&&\times \left[ \cos \zeta _{+}\sum_{k=0}^{3}L_{k}^{SS}\cos ^{k}\chi
_{p}\right.  \notag \\
&&\left. +\sin \zeta _{+}\sin \chi _{p}\sum_{k=0}^{2}K_{k}^{SS}\cos ^{k}\chi
_{p}\right] ~,  \label{lrchiSS}
\end{eqnarray}%
\begin{eqnarray}
\mathfrak{l}_{rQM}\left( \chi _{p}\right) &=&\frac{\eta }{2}%
\sum_{i=1}^{2}\chi _{i}^{2}w_{i}\nu ^{2i-3}\sin ^{2}\kappa _{i}  \notag \\
&&\times \left[ \cos 2\zeta _{i}\sum_{k=0}^{3}L_{k}^{QM}\cos ^{k}\chi
_{p}\right.  \notag \\
&&\left. +\sin 2\zeta _{i}\sin \chi _{p}\sum_{k=0}^{2}K_{k}^{QM}\cos
^{k}\chi _{p}\right] ~,  \label{lrchiQM}
\end{eqnarray}%
where the coefficients $L_{kl}^{2PN}$, $L_{k}^{SS}$, $K_{k}^{SS}$, $%
L_{k}^{QM}$, and $K_{k}^{QM}$ are enlisted in Table \ref{tablelrchip}. 
\begin{table}[tbp]
\caption{The coefficients of $\mathfrak{l}_{r}\left( \protect\chi %
_{p}\right) $ in Eqs. (\protect\ref{lrchi2PN}), (\protect\ref{lrchiSS}), and
(\protect\ref{lrchiQM}).}
\label{tablelrchip}
\begin{center}
\begin{tabular}{cc}
\hline\hline
$Coefficient$ & $Expression$ \\ \hline\hline
$L_{00}^{2PN}$ & $\frac{1}{96}\left[ 432e_{r0}^{3}\eta +e_{r0}^{2}\left(
-117\eta ^{2}+54\eta +48\right) \right. $ \\ 
& $+32e_{r0}\left( 2\eta ^{2}-83\eta +50\right)$ \\ 
& $\left. -48\left( \eta ^{2}-5\eta +6\right) \right]$ \\ \hline
$L_{10}^{2PN}$ & $\frac{e_{r0}}{8}\left[ \left( 2\eta -33\right) \eta
e_{r0}^{2}\right. $ \\ 
& $\left. -116\eta ^{2}+256\eta -160\right]$ \\ \hline
$L_{20}^{2PN}$ & $\frac{1}{8}\left[ e_{r0}^{2}\left( 9\eta ^{2}-3\eta
-4\right) \right.$ \\ 
& $\left. +4\left( \eta ^{2}-5\eta +6\right) \right]$ \\ \hline
$L_{30}^{2PN}$ & $\frac{e_{r0}}{24}\left[ -3\left( 2\eta +3\right) \eta
e_{r0}^{2}\right.$ \\ 
& $\left. +32\eta ^{2}-104\eta +80\right]$ \\ \hline
$L_{40}^{2PN}$ & $\frac{3e_{r0}^{2}(\eta -2)\eta }{32}$ \\ \hline
$L_{01}^{2PN}$ & $\frac{1}{8}\left[ e_{r0}^{2}\left( -9\eta ^{2}+3\eta
+4\right) \right. $ \\ 
& $\left. -4\left( \eta ^{2}-5\eta +6\right) \right] $ \\ \hline
$L_{11}^{2PN}$ & $\frac{e_{r0}}{8}\left[ 3\left( 2\eta +3\right) \eta
e_{r0}^{2}\right. $ \\ 
& $\left. -32\eta ^{2}+104\eta -80\right] $ \\ \hline
$L_{21}^{2PN}$ & $-\frac{9e_{r0}^{2}(\eta -2)\eta }{16} $ \\ \hline
$L_{31}^{2PN}$ & $0 $ \\ \hline
$L_{41}^{2PN}$ & $0 $ \\ \hline
$L_{02}^{2PN}$ & $\frac{3e_{r0}^{2}(\eta -2)\eta }{32} $ \\ \hline
$L_{12}^{2PN}$ & $0 $ \\ \hline
$L_{22}^{2PN}$ & $0$ \\ \hline
$L_{32}^{2PN}$ & $0$ \\ \hline
$L_{42}^{2PN}$ & $0$ \\ \hline\hline
$L_{0}^{SS}$ & $2e_{r0}+3 $ \\ \hline
$L_{1}^{SS}$ & $0 $ \\ \hline
$L_{2}^{SS}$ & $-3 $ \\ \hline
$L_{3}^{SS}$ & $-2e_{r0} $ \\ \hline
$K_{0}^{SS}$ & $-e_{r0} $ \\ \hline
$K_{1}^{SS}$ & $-3 $ \\ \hline
$K_{2}^{SS}$ & $-2e_{r0}$ \\ \hline\hline
$L_{0}^{QM}$ & $-\left( 2e_{r0}+3\right) $ \\ \hline
$L_{1}^{QM}$ & $0 $ \\ \hline
$L_{2}^{QM}$ & $-3 $ \\ \hline
$L_{3}^{QM}$ & $-2e_{r0} $ \\ \hline
$K_{0}^{QM}$ & $-e_{r0} $ \\ \hline
$K_{1}^{QM}$ & $-3 $ \\ \hline
$K_{2}^{QM}$ & $-2e_{r0} $ \\ \hline\hline
\end{tabular}%
\end{center}
\end{table}

\subsection{$\protect\chi _{p}$ dependence of $e_{r}$\label{erevo}}

From $\dot{e}_{r}$ and $\dot{\chi}_{p}$ given by Eqs. (37) and (43) of Ref. 
\cite{chameleon}, respectively, we find%
\begin{eqnarray}
e_{r}\left( \chi _{p}\right) &=&e_{r0}+\int_{0}^{\chi _{p}}\frac{\dot{e}_{r}%
}{\dot{\chi}_{p}}d\chi _{p}~=  \notag \\
&=&e_{r0}+\frac{1}{\mathfrak{l}_{r0}^{2}}e_{rPN}\left( \chi _{p}\right) +%
\frac{1}{\mathfrak{l}_{r0}^{3}}e_{rSO}\left( \chi _{p}\right)  \notag \\
&&+\frac{1}{\mathfrak{l}_{r0}^{4}}e_{rQM}\left( \chi _{p}\right) +\frac{1}{%
\mathfrak{l}_{r0}^{4}}e_{rSS}\left( \chi _{p}\right)  \notag \\
&&+\frac{1}{\mathfrak{l}_{r0}^{4}}e_{r2PN}\left( \chi _{p}\right) ~,
\end{eqnarray}%
with%
\begin{equation}
e_{rPN}\left( \chi _{p}\right) =\sum_{k=0}^{3}E_{k}^{PN}\cos ^{k}\chi _{p}~,
\label{erchiPN}
\end{equation}%
\begin{eqnarray}
e_{rSO}\left( \chi _{p}\right) &=&\frac{\eta }{2}\left( 1-e_{r0}^{2}\right)
\left( 1-\cos \chi _{p}\right) \times  \notag \\
&&\times \sum_{i=1}^{2}\left( 4^{2k-3}+3\right) \chi _{i}\cos \kappa _{i}~,
\end{eqnarray}%
\begin{equation}
e_{r2PN}\left( \chi _{p}\right)
=\sum_{l=0}^{3}\sum_{k=0}^{6}E_{kl}^{2PN}\cos ^{k}\chi _{p}\sin ^{2l}\chi
_{p}~,  \label{erchi2PN}
\end{equation}%
\begin{gather}
e_{rSS}\left( \chi _{p}\right) =\eta \chi _{1}\chi _{2}\left[
\sum_{k=0}^{5}E_{k}^{SS}\cos ^{k}\chi _{p}\right.  \notag \\
\left. +\sin \kappa _{1}\sin \kappa _{2}\sin \zeta _{+}\sin \chi
_{p}\sum_{k=0}^{4}F_{k}^{SS}\cos ^{k}\chi _{p}\right] ~,  \label{erchiSS}
\end{gather}%
\begin{eqnarray}
e_{rQM}\left( \chi _{p}\right) &=&\frac{\eta }{2}\sum_{i=1}^{2}\chi
_{i}^{2}\nu ^{2i-3}w_{i}  \notag \\
&&\times \left[ \sum_{k=0}^{6}E_{k}^{QM}\cos ^{k}\chi _{p}\right.  \notag \\
&&+\sin 2\zeta _{i}\sin ^{2}\kappa _{i}\sin \chi _{p}  \notag \\
&&\left. \times \sum_{k=0}^{4}F_{k}^{QM}\cos ^{k}\chi _{p}\right] ~.
\label{erchiQM}
\end{eqnarray}%
The coefficients $E_{kl}^{2PN}$, $E_{k}^{SS}$, $F_{k}^{SS}$, $E_{k}^{QM}$,
and $F_{k}^{QM}~$of $e_{r}\left( \chi _{p}\right) $ for the PN, SS,\ and QM
contributions are collected in Table \ref{tableerchippnssqm}. The
coefficients $E_{k}^{2PN}$ and $F_{k}^{2PN}$ are enlisted in Table \ref%
{tableerchip2pn}. 
\begin{table}[tbp]
\caption{The coefficients of $e_{r}\left( \protect\chi _{p}\right) $ in Eqs.
(\protect\ref{erchiPN}), (\protect\ref{erchiSS}), and (\protect\ref{erchiQM}%
). }
\label{tableerchippnssqm}
\begin{center}
\begin{tabular}{cc}
\hline\hline
$Coefficient$ & $Expression$ \\ \hline\hline
$E_{0}^{PN}$ & $3-\eta +\left( 5-4\eta \right) e_{r0}+e_{r0}^{2}\left(
7-6\eta \right)$ \\ \hline
$E_{1}^{PN}$ & $-\left[ 3-\eta +e_{r0}^{2}\left( 7-\frac{11}{2}\eta \right) %
\right] $ \\ \hline
$E_{2}^{PN}$ & $-\left( 5-4\eta \right) e_{r0}$ \\ \hline
$E_{3}^{PN}$ & $\frac{\eta }{2}e_{r0}^{2}$ \\ \hline\hline
$E_{0}^{SS}$ & $-\cos \kappa _{1}\cos \kappa _{2}\left(
e_{r0}^{2}+3e_{r0}+3\right)$ \\ 
& $+\frac{1}{2}\sin \kappa _{1}\sin \kappa _{2}\left[ \left(
e_{r0}^{2}+3e_{r0}+3\right) \cos \zeta _{-}\right.$ \\ 
& $+\left. \left( 7e_{r0}^{2}+15e_{r0}+5\right) \cos \zeta _{+}\right]$ \\ 
\hline
$E_{1}^{SS}$ & $\frac{3}{2}\sin \kappa _{1}\sin \kappa _{2}\left( 3\cos
\zeta _{+}-\cos \zeta _{-}\right) $ \\ 
& $+3\cos \kappa _{1}\cos \kappa _{2}$ \\ \hline
$E_{2}^{SS}$ & $\frac{3}{2}e_{r0}\sin \kappa _{1}\sin \kappa _{2}\left( \cos
\zeta _{+}-\cos \zeta _{-}\right) $ \\ 
& $+3e_{r0}\cos \kappa _{1}\cos \kappa _{2}$ \\ \hline
$E_{3}^{SS}$ & $-\frac{1}{2}e_{r0}^{2}\sin \kappa _{1}\sin \kappa _{2}\left(
\cos \zeta _{+}+\cos \zeta _{-}\right) $ \\ 
& $-7\sin \kappa _{1}\sin \kappa _{2}\cos \zeta _{+} $ \\ 
& $+e_{r0}^{2}\cos \kappa _{1}\cos \kappa _{2} $ \\ \hline
$E_{4}^{SS}$ & $-9e_{r0}\sin \kappa _{1}\sin \kappa _{2}\cos \zeta _{+} $ \\ 
\hline
$E_{5}^{SS}$ & $-3e_{r0}^{2}\sin \kappa _{1}\sin \kappa _{2}\cos \zeta _{+}$
\\ \hline
$F_{0}^{SS}$ & $1-e_{r0}^{2}$ \\ \hline
$F_{1}^{SS}$ & $-3e_{r0}$ \\ \hline
$F_{2}^{SS}$ & $-\left( 2e_{r0}^{2}+7\right)$ \\ \hline
$F_{3}^{SS}$ & $-9e_{r0}$ \\ \hline
$F_{4}^{SS}$ & $-3e_{r0}^{2}$ \\ \hline\hline
$E_{0}^{QM}$ & $\left( 7e_{r0}^{2}+15e_{r0}+5\right) \cos ^{2}\zeta _{i}\sin
^{2}\kappa _{i}$ \\ 
& $+\left( 2e_{r0}^{2}+3e_{r0}-2\right) \cos ^{2}\kappa _{i}$ \\ 
& $-3e_{r0}^{2}-6e_{r0}-1 $ \\ \hline
$E_{1}^{QM}$ & $9\cos ^{2}\kappa _{i}\sin ^{2}\zeta _{i}+3\left( 3\cos
^{2}\zeta _{i}-2\right)$ \\ \hline
$E_{2}^{QM}$ & $3e_{r0}\left( 2-\cos ^{2}\zeta _{i}\right) \cos ^{2}\kappa
_{i}$ \\ 
& $-3e_{r0}\sin ^{2}\zeta _{i}$ \\ \hline
$E_{3}^{QM}$ & $e_{r0}^{2}\cos ^{2}\kappa _{i}$ \\ 
& $+\left[ -\left( e_{r0}^{2}+14\right) \cos ^{2}\zeta _{i}+7\right] \sin
^{2}\kappa _{i}$ \\ \hline
$E_{4}^{QM}$ & $-9e_{r0}\sin ^{2}\kappa _{i}\cos 2\zeta _{i} $ \\ \hline
$E_{5}^{QM}$ & $-3e_{r0}^{2}\sin ^{2}\kappa _{i}\cos 2\zeta _{i}$ \\ \hline
$F_{0}^{QM}$ & $\left( 1-e_{r0}^{2}\right) $ \\ \hline
$F_{1}^{QM}$ & $-3e_{r0}$ \\ \hline
$F_{2}^{QM}$ & $-\left( 2e_{r0}^{2}+7\right)$ \\ \hline
$F_{3}^{QM}$ & $-9e_{r0} $ \\ \hline
$F_{4}^{QM}$ & $-3e_{r0}^{2}$ \\ \hline\hline
\end{tabular}%
\end{center}
\end{table}
\begin{table}[th]
\caption{The coefficients of $e_{r}\left( \protect\chi _{p}\right) $ in Eq. (%
\protect\ref{erchi2PN}).}
\label{tableerchip2pn}
\begin{center}
\begin{tabular}{cc}
\hline\hline
$Coefficient$ & $Expression$ \\ \hline\hline
$E_{00}^{2PN}$ & $\frac{1}{1920e_{r0}}\left[ 1920e_{r0}^{5}\eta (3\eta
+8)\right. $ \\ 
& $+e_{r0}^{4}\left( -1845\eta ^{2}+8880\eta +1800\right) $ \\ 
& $+32e_{r0}^{3}\left( 232\eta ^{2}-2825\eta +1740\right)$ \\ 
& $-180e_{r0}^{2}\left( 29\eta ^{2}+89\eta +60\right)$ \\ 
& $+160e_{r0}\left( 8\eta ^{2}-187\eta -60\right)$ \\ 
& $\left. -480(\eta -3)^{2}\right] $ \\ \hline
$E_{10}^{2PN}$ & $-\frac{1}{64}\left[ e_{r0}^{4}\eta \left( 161\eta
+477\right) \right. $ \\ 
& $+4e_{r0}^{2}\left( 136\eta ^{2}-849\eta +564\right) $ \\ 
& $\left. +16\eta \left( 8\eta -85\right) \right] $ \\ \hline
$E_{20}^{2PN}$ & $\frac{1}{256e_{r0}}\left[ e_{r0}^{4}\left( 269\eta
^{2}-1312\eta -256\right) \right. $ \\ 
& $+32e_{r0}^{2}\left( 5\eta ^{2}+109\eta +20\right)$ \\ 
& $\left. +64\left( \eta -3\right) ^{2}\right] $ \\ \hline
$E_{30}^{2PN} $ & $\frac{1}{384}\left[ -3e_{r0}^{4}\eta \left( 53\eta
+73\right) \right. $ \\ 
& $+8e_{r0}^{2}\left( 208\eta ^{2}-269\eta +300\right)$ \\ 
& $\left. +128\left( 4\eta ^{2}-17\eta +15\right) \right]$ \\ \hline
$E_{40}^{2PN}$ & $\frac{e_{r0}}{128}\left[ e_{r0}^{2}\left( -13\eta
^{2}+64\eta +8\right) \right.$ \\ 
& $\left. +268\eta ^{2}-676\eta +400\right]$ \\ \hline
$E_{50}^{2PN}$ & $-\frac{3e_{r0}^{2}\eta }{640}\left[ 5e_{r0}^{2}(3\eta
-1)-64\eta +80\right] $ \\ \hline
$E_{60}^{2PN}$ & $\frac{3e_{r0}^{3}\eta ^{2}}{256}$ \\ \hline
$E_{01}^{2PN}$ & $-E_{20}^{2PN}$ \\ \hline
$E_{11}^{2PN}$ & $-3E_{30}^{2PN}$ \\ \hline
$E_{21}^{2PN}$ & $-\frac{3}{2}E_{40}^{2PN}$ \\ \hline
$E_{31}^{2PN}$ & $-\frac{1}{10}E_{50}^{2PN}$ \\ \hline
$E_{41}^{2PN}$ & $-9E_{60}^{2PN}$ \\ \hline
$E_{51}^{2PN}$ & $0$ \\ \hline
$E_{61}^{2PN}$ & $0$ \\ \hline
$E_{02}^{2PN}$ & $E_{40}^{2PN}$ \\ \hline
$E_{12}^{2PN}$ & $\frac{1}{5}E_{50}^{2PN} $ \\ \hline
$E_{22}^{2PN}$ & $9E_{60}^{2PN}$ \\ \hline
$E_{32}^{2PN}$ & $0$ \\ \hline
$E_{42}^{2PN}$ & $0$ \\ \hline
$E_{52}^{2PN}$ & $0$ \\ \hline
$E_{62}^{2PN}$ & $0$ \\ \hline
$E_{03}^{2PN}$ & $-E_{60}^{2PN}$ \\ \hline
$E_{13}^{2PN}$ & $0$ \\ \hline
$E_{23}^{2PN}$ & $0$ \\ \hline
$E_{33}^{2PN}$ & $0$ \\ \hline
$E_{43}^{2PN}$ & $0$ \\ \hline
$E_{53}^{2PN}$ & $0$ \\ \hline
$E_{63}^{2PN}$ & $0$ \\ \hline\hline
\end{tabular}%
\end{center}
\end{table}

\subsection{Dimensionless 2PN orbital period}

We insert the expressions of $\mathfrak{l}_{r}\left( \chi _{p}\right) $\ and 
$e_{r}\left( \chi _{p}\right) $ into the integral (\ref{perioddef}) and
expand it in Taylor series to 2PN order.

The integral (\ref{perioddef}) leads to the PN expansion:%
\begin{eqnarray}
\mathfrak{T} &=&\mathfrak{T}_{0}\left( 1+\frac{\tau _{0PN}}{\mathfrak{l}%
_{r}^{2}}+\frac{\tau _{0SO}}{\mathfrak{l}_{r}^{3}}+\frac{\tau _{0SS}}{%
\mathfrak{l}_{r}^{4}}\right.  \notag \\
&&\left. +\frac{\tau _{0QM}}{\mathfrak{l}_{r}^{4}}+\frac{\tau _{02PN}}{%
\mathfrak{l}_{r}^{4}}\right) ~,  \label{period}
\end{eqnarray}%
where the lower index $0$ stands for $\chi _{p}=0$. As explained in the main
text $\tau _{0PN}/\mathfrak{l}_{r}^{2}$ and $\tau _{02PN}/\mathfrak{l}%
_{r}^{4}$ give PN- and 2PN-order contributions, respectively.

The terms $\mathfrak{T}_{0}$, $\tau _{0PN}$, $\tau _{0SO}$, $\tau _{0SS}$, $%
\tau _{0QM}$ and $\tau _{02PN}$ are found by exploring the expressions $%
\mathfrak{l}_{r}\left( \chi _{p}\right) $ and $e_{r}\left( \chi _{p}\right) $
derived above. They read as%
\begin{equation}
\mathfrak{T}_{0}=\frac{2\pi \mathfrak{l}_{r0}^{3}}{\left(
1-e_{r0}^{2}\right) ^{3/2}}~,  \label{tau0expr}
\end{equation}%
\begin{equation*}
\tau _{0PN}=-\frac{\left( 1-e_{r0}^{2}\right) \left( e_{r0}^{2}(7\eta
-6)+2e_{r0}(5\eta -3)+4\eta -18\right) }{2(e_{r0}-1)^{2}}~,
\end{equation*}%
\begin{equation}
\tau _{0SO}=0~,
\end{equation}%
\begin{eqnarray}
\tau _{02PN} &=&\frac{1}{40(1-e_{r0})^{2}e_{r0}^{4}}\left[
\sum_{k=0}^{10}U_{k}e_{r0}^{k}\right.  \notag \\
&&\left. -\frac{\left( 1-e_{r0}^{2}\right) ^{3/2}}{2(1-e_{r0})}%
\sum_{k=0}^{7}V_{k}e_{r0}^{k}\right] ~,  \label{tau02PN}
\end{eqnarray}%
\begin{eqnarray}
\tau _{0SS} &=&-\frac{3\chi _{1}\chi _{2}\left( 1+e_{r0}\right) ^{2}\eta }{%
\left( 1-e_{r0}\right) \mathfrak{l}_{r0}^{4}}\left[ \cos \kappa _{1}\cos
\kappa _{2}+\sin \kappa _{1}\sin \kappa _{2}\right.  \notag \\
&&\left. \times \left( \sin \zeta _{1}\sin \zeta _{2}-2\cos \zeta _{1}\cos
\zeta _{2}\right) \right] ~,
\end{eqnarray}%
\begin{eqnarray}
\tau _{0QM} &=&-\frac{3\eta \left( 1+e_{r0}\right) ^{2}}{2\left(
1-e_{r0}\right) \mathfrak{l}_{r0}^{4}}\sum_{i=1}^{2}\chi _{i}^{2}\nu
^{2i-3}w_{i}  \notag \\
&&\times \left( 1-3\sin ^{2}\kappa _{i}\cos ^{2}\zeta _{i}\right) ~.
\end{eqnarray}%
The coefficients $U_{k}$ and $V_{k}$ of $\tau _{02PN}$ are enlisted in Table %
\ref{tabletau02pn}. 
\begin{table}[tbp]
\caption{The coefficients $U_{k}$ and $V_{k}$ of $\protect\tau _{02PN}$ in
Eq. (\protect\ref{tau02PN}).}
\label{tabletau02pn}
\begin{center}
\begin{tabular}{cc}
\hline\hline
$Coefficient$ & $Expression$ \\ \hline\hline
$U_{10}$ & $-105\left( \eta +1\right) \eta$ \\ \hline
$U_{9}$ & $10\left( -559\eta +297\eta ^{2}+228\right) $ \\ \hline
$U_{8}$ & $5\left( -2674\eta +1289\eta ^{2}+1336\right) $ \\ \hline
$U_{7}$ & $-4\left( -235\eta +186\eta ^{2}-280\right)$ \\ \hline
$U_{6}$ & $-482\eta ^{2}+25\eta +1120$ \\ \hline
$U_{5}$ & $2\left( -17\,245\eta +5496\eta ^{2}+12\,200\right)$ \\ \hline
$U_{4}$ & $2\left( -2240\eta +79\eta ^{2}+3350\right)$ \\ \hline
$U_{3}$ & $-4\left( -7075\eta +2137\eta ^{2}+5300\right)$ \\ \hline
$U_{2}$ & $2\left( -3915\eta +1317\eta ^{2}+2050\right) $ \\ \hline
$U_{1}$ & $40\left( -254\eta +67\eta ^{2}+210\right)$ \\ \hline
$U_{0}$ & $-20\left( -238\eta +65\eta ^{2}+180\right)$ \\ \hline\hline
$V_{7}$ & $5\left( -932\eta +509\eta ^{2}+504\right)$ \\ \hline
$V_{6}$ & $-5\left( -2700\eta +1399\eta ^{2}+1352\right) $ \\ \hline
$V_{5}$ & $1427\eta ^{2}+920\eta -4240 $ \\ \hline
$V_{4}$ & $12079\eta ^{2}-35880\eta +25680 $ \\ \hline
$V_{3}$ & $-4\left( -6875\eta +2606\eta ^{2}+3650\right) $ \\ \hline
$V_{2}$ & $-4\left( -4735\eta +998\eta ^{2}+4850\right)$ \\ \hline
$V_{1}$ & $40\left( -746\eta +199\eta ^{2}+600\right) $ \\ \hline
$V_{0}$ & $-40\left( -238\eta +65\eta ^{2}+180\right) $ \\ \hline\hline
\end{tabular}%
\end{center}
\end{table}

\subsection{Secular shape variables $\mathfrak{\bar{l}}_{r}$ and $\bar{e}%
_{r} $\label{lreraver}}

Next we calculate the time averages of the shape variables $\mathfrak{l}_{r}$
and $e_{r}$ during one radial period 
\begin{eqnarray}
\mathfrak{\bar{l}}_{r} &=&\frac{1}{\mathfrak{T}}\int_{0}^{2\pi }\frac{%
\mathfrak{l}_{r}\left( \chi _{p}\right) }{\dot{\chi}_{p}}d\chi _{p}~,
\label{lraverdefkif} \\
\bar{e}_{r} &=&\frac{1}{\mathfrak{T}}\int_{0}^{2\pi }\frac{e_{r}\left( \chi
_{p}\right) }{\dot{\chi}_{p}}d\chi _{p}~,  \label{eraverdefkif}
\end{eqnarray}

in terms of their initial values at $\chi _{p}=0$, with the decomposition of 
$\mathfrak{T}$ given in Eq. (\ref{period}). Their PN expansion can be
formally written as

\begin{eqnarray}
\mathfrak{\bar{l}}_{r} &\!=\!&\frac{\mathfrak{l}_{r0N}\!+\!\mathfrak{\bar{l}}%
_{rPN}\!+\!\mathfrak{\bar{l}}_{rSO}\!+\!\mathfrak{\bar{l}}_{rSS}\!+\!%
\mathfrak{\bar{l}}_{rQM}\!+\!\mathfrak{\bar{l}}_{r2PN}}{\mathfrak{T}}~,
\label{lraverexpr} \\
\bar{e}_{r} &\!=\!&\frac{e_{r0N}\!+\!\bar{e}_{rPN}\!+\!\bar{e}_{rSO}\!+\!%
\bar{e}_{rSS}\!+\!\bar{e}_{rQM}\!+\!\bar{e}_{r2PN}}{\mathfrak{T}}~.
\label{eraverexpr}
\end{eqnarray}

The contributions to the integral in Eq. (\ref{lraverdefkif}) become 
\begin{equation}
\mathfrak{\bar{l}}_{rN}=\frac{2\pi \mathfrak{l}_{r0}^{4}}{\left(
1-e_{r0}^{2}\right) ^{3/2}}~,  \label{lravernewtexpr}
\end{equation}%
\begin{eqnarray}
\mathfrak{\bar{l}}_{rPN} &=&-\frac{\pi \mathfrak{l}_{r0}^{2}}{(1-e_{r0})^{2}%
\sqrt{1-e_{r0}^{2}}}\left[ e_{r0}^{2}(3\eta +2)\right.  \notag \\
&&\left. +14e_{r0}(\eta -1)+4\eta -18\right] ~,
\end{eqnarray}%
\begin{eqnarray}
\mathfrak{\bar{l}}_{rSO} &=&-\frac{\pi e_{r0}\eta \mathfrak{l}_{r0}}{%
(1-e_{r0})\sqrt{1-e_{r0}^{2}}}  \notag \\
&&\times \sum_{k=1}^{2}\left( 4^{2k-3}+3\right) \chi _{k}\cos \kappa _{k}~,
\end{eqnarray}%
\begin{eqnarray}
\mathfrak{\bar{l}}_{r2PN} &=&\frac{\pi }{120e_{r0}^{4}}\sum_{k=0}^{4}\bar{L}%
_{k}^{2PN}e_{r0}^{k}  \notag \\
&&-\frac{\pi \sqrt{1-e_{r0}^{2}}}{60(e_{r0}-1)^{4}e_{r0}^{4}}\sum_{k=0}^{8}%
\bar{K}_{k}^{2PN}e_{r0}^{k}~,  \label{lraver2PN}
\end{eqnarray}%
\begin{eqnarray}
\mathfrak{\bar{l}}_{rSS} &=&\frac{\pi \chi _{1}\chi _{2}\eta }{16\left(
1-e_{r0}\right) e_{r0}^{2}\left( 1-e_{r0}^{2}\right) ^{3/2}}  \notag \\
&&\times \left( \cos \kappa _{1}\cos \kappa _{2}\bar{L}^{SS}\right.  \notag
\\
&&\left. +\sin \kappa _{1}\sin \kappa _{2}\bar{K}^{SS}\right) ~,
\label{lraverSS}
\end{eqnarray}%
\begin{eqnarray}
\mathfrak{\bar{l}}_{rQM} &=&\frac{\pi \left( e_{r0}^{2}+2\right) \eta }{%
256e_{r0}\left( 1-e_{r0}^{2}\right) ^{5/2}}\sum_{k=1}^{2}\chi _{k}^{2}\nu
^{2k-3}w_{k}  \notag \\
&&\times \left[ -4\left( 47e_{r0}^{3}+1050e_{r0}^{2}+488e_{r0}\right. \right.
\notag \\
&&\left. +480\right) \sin ^{2}\kappa _{k}\cos 2\zeta _{k}  \notag \\
&&+16\left( 5e_{r0}^{3}+21e_{r0}^{2}+15e_{r0}+6\right)  \notag \\
&&\left. \times \left( 3\cos 2\kappa _{k}+1\right) \right] ~.
\end{eqnarray}%
The coefficients $\bar{L}_{k}^{2PN}$, $\bar{K}_{k}^{2PN}$, $\bar{L}^{SS}$,
and $\bar{K}^{SS}$ are collected in Table \ref{tablelraver}. 
\begin{table}[tbp]
\caption{The coefficients of $\mathfrak{\bar{l}}_{r}$ in Eqs. (\protect\ref%
{lraver2PN}) and (\protect\ref{lraverSS}).}
\label{tablelraver}
\begin{center}
\begin{tabular}{cc}
\hline\hline
$Coefficient$ & $Expression$ \\ \hline\hline
$\bar{L}_{4}^{2PN} $ & $15\left( 467\eta ^{2}-580\eta +296\right)$ \\ \hline
$\bar{L}_{3}^{2PN}$ & $480\left( 4\eta ^{2}-3\eta +5\right)$ \\ \hline
$\bar{L}_{2}^{2PN}$ & $-4\left( 3001\eta ^{2}-9445\eta +6610\right)$ \\ 
\hline
$\bar{L}_{1}^{2PN}$ & $-480\left( \eta ^{2}-8\eta +15\right) $ \\ \hline
$\bar{L}_{0}^{2PN}$ & $120\left( 65\eta ^{2}-238\eta +180\right)$ \\ \hline
$\bar{K}_{8}^{2PN}$ & $15\eta (29\eta -3)$ \\ \hline
$\bar{K}_{7}^{2PN}$ & $-60\left( 129\eta ^{2}-188\eta +74\right)$ \\ \hline
$\bar{K}_{6}^{2PN}$ & $-15\left( 116\eta ^{2}-711\eta +304\right)$ \\ \hline
$\bar{K}_{5}^{2PN}$ & $2\left( 5516\eta ^{2}-15155\eta +5420\right) $ \\ 
\hline
$\bar{K}_{4}^{2PN}$ & $-4\left( 5347\eta ^{2}-13720\eta +6400\right) $ \\ 
\hline
$\bar{K}_{3}^{2PN}$ & $8\left( \eta ^{2}+2735\eta -4265\right)$ \\ \hline
$\bar{K}_{2}^{2PN}$ & $20308\eta ^{2}-81610\eta +71380 $ \\ \hline
$\bar{K}_{1}^{2PN}$ & $-720\left( 22\eta ^{2}-82\eta +65\right)$ \\ \hline
$\bar{K}_{0}^{2PN}$ & $60\left( 65\eta ^{2}-238\eta +180\right)$ \\ 
\hline\hline
$\bar{L}^{SS}$ & $4e_{r0}\left( 4e_{r0}^{4}+29e_{r0}^{3}\right.$ \\ 
& $\left. +30e_{r0}^{2}+48e_{r0}+24\right)$ \\ \hline
$\bar{K}^{SS}$ & $\cos \left( \zeta _{1}+\zeta _{2}\right) \left[
96e_{r0}^{3}-236e_{r0}^{2}\right.$ \\ 
& $-171e_{r0}^{4}+95e_{r0}^{5}+56e_{r0}-32 $ \\ 
& $\left. -32\sqrt{1-e_{r0}^{2}}\left( e_{r0}^{2}-e_{r0}^{3}+e_{r0}-1\right) %
\right]$ \\ 
& $-2e_{r0}\cos \left( \zeta _{1}-\zeta _{2}\right) \left(
4e_{r0}^{4}+29e_{r0}^{3}\right. $ \\ 
& $\left. +30e_{r0}^{2}+48e_{r0}+24\right)$ \\ \hline\hline
\end{tabular}%
\end{center}
\end{table}

The integral of Eq. (\ref{eraverdefkif}) results in%
\begin{equation}
\bar{e}_{rN}=\frac{2\pi e_{r0}\mathfrak{l}_{r0}^{3}}{\left(
1-e_{r0}^{2}\right) ^{3/2}}~,  \label{eravernewtexpr}
\end{equation}%
\begin{eqnarray}
\bar{e}_{rPN} &=&\frac{\pi \mathfrak{l}_{r0}}{e_{r0}}\left\{ 2(3\eta
-5)\right.  \notag \\
&&+\frac{\sqrt{1-e_{r0}^{2}}}{(1-e_{r0})^{3}(e_{r0}+1)}\left[
4e_{r0}^{4}\left( \eta -2\right) \right.  \notag \\
&&-20e_{r0}^{3}\left( \eta -1\right) +e_{r0}^{2}\left( 22-9\eta \right) 
\notag \\
&&\left. \left. +2e_{r0}\left( 5\eta -7\right) -6\eta +10\right] \right\} ~,
\end{eqnarray}%
\begin{eqnarray}
\bar{e}_{rSO} &=&\frac{\pi (e_{r0}+1)\eta }{\sqrt{1-e_{r0}^{2}}}  \notag \\
&&\times \sum_{k=1}^{2}\left( 4^{2k-3}+3\right) \chi _{k}\cos \kappa _{k}~,
\end{eqnarray}%
\begin{eqnarray}
\bar{e}_{r2PN} &=&\frac{\pi }{480e_{r0}^{3}\mathfrak{l}_{r0}}\sum_{k=0}^{4}%
\bar{E}_{k}^{2PN}e_{r0}^{k}  \notag \\
&&-\frac{\pi \sqrt{1-e_{r0}^{2}}}{60(e_{r0}-1)^{4}e_{r0}^{3}\mathfrak{l}_{r0}%
}\sum_{k=0}^{8}\bar{F}_{k}^{2PN}e_{r0}^{k}~,  \label{eraver2PN}
\end{eqnarray}%
\begin{eqnarray}
\bar{e}_{rSS} &=&\frac{3\pi \chi _{1}\chi _{2}\eta }{16\left(
1-e_{r0}\right) e_{r0}^{3}\left( 1-e_{r0}^{2}\right) ^{2}\mathfrak{l}_{r0}} 
\notag \\
&&\times \left( \cos \kappa _{1}\cos \kappa _{2}\bar{E}^{SS}\right.  \notag
\\
&&\left. +\sin \kappa _{1}\sin \kappa _{2}\bar{F}^{SS}\right) ~,
\label{eraverSS}
\end{eqnarray}%
\begin{eqnarray}
\bar{e}_{rQM} &=&\frac{\pi \left( e_{r0}^{2}+2\right) \eta }{128\left(
1-e_{r0}^{2}\right) ^{5/2}\mathfrak{l}_{r0}}\sum_{k=1}^{2}\chi _{k}^{2}\nu
^{2k-3}w_{k}  \notag \\
&&\times \left[ -4\left( 10e_{r0}^{3}+609e_{r0}^{2}+260e_{r0}\right. \right.
\notag \\
&&\left. +336\right) \sin ^{2}\kappa _{k}\cos 2\zeta _{k}+8\left(
4e_{r0}^{3}\right.  \notag \\
&&\left. +27e_{r0}^{2}+15e_{r0}+12\right)  \notag \\
&&\left. \times \left( 3\cos 2\kappa _{k}+1\right) \right] ~.
\end{eqnarray}%
The coefficients $\bar{E}_{k}^{2PN}$, $\bar{F}_{k}^{2PN}$, $\bar{E}^{SS}$,
and $\bar{F}^{SS}$ are collected in Table \ref{tableeraver}. 
\begin{table}[tbp]
\caption{The coefficients of $\bar{e}_{r}$ in Eqs. (\protect\ref{eraver2PN})
and (\protect\ref{eraverSS}).}
\label{tableeraver}
\begin{center}
\begin{tabular}{cc}
\hline\hline
$Coefficient$ & $Expression$ \\ \hline\hline
$\bar{E}_{4}^{2PN} $ & $15\left( 1111\eta ^{2}-1624\eta +528\right)$ \\ 
\hline
$\bar{E}_{3}^{2PN}$ & $4800\left( 4\eta ^{2}-7\eta +4\right)$ \\ \hline
$\bar{E}_{2}^{2PN}$ & $-4\left( 151\eta ^{2}-7550\eta +5640\right) $ \\ 
\hline
$\bar{E}_{1}^{2PN}$ & $2880(\eta -3)\eta $ \\ \hline
$\bar{E}_{0}^{2PN}$ & $8\left( 1501\eta ^{2}-8090\eta +7260\right)$ \\ \hline
$\bar{F}_{8}^{2PN}$ & $120(\eta -3)\eta$ \\ \hline
$\bar{F}_{7}^{2PN}$ & $-60\left( 44\eta ^{2}-107\eta +44\right) $ \\ \hline
$\bar{F}_{6}^{2PN}$ & $-15\left( 317\eta ^{2}-537\eta +96\right)$ \\ \hline
$\bar{F}_{5}^{2PN}$ & $4\left( 457\eta ^{2}-1430\eta -960\right)$ \\ \hline
$\bar{F}_{4}^{2PN}$ & $-5362\eta ^{2}+18335\eta -4380 $ \\ \hline
$\bar{F}_{3}^{2PN}$ & $-6\left( 714\eta ^{2}-4315\eta +5420\right)$ \\ \hline
$\bar{F}_{2}^{2PN}$ & $21\left( 391\eta ^{2}-2110\eta +2100\right)$ \\ \hline
$\bar{F}_{1}^{2PN}$ & $-4\left( 1411\eta ^{2}-7820\eta +7260\right)$ \\ 
\hline
$\bar{F}_{0}^{2PN}$ & $1501\eta ^{2}-8090\eta +7260$ \\ \hline\hline
$\bar{E}^{SS}$ & $16e_{r0}^{3}(1+e_{r0})^{2}\sqrt{1-e_{r0}^{2}}\left(
e_{r0}^{2}+2\right)$ \\ \hline
$\bar{F}^{SS}$ & $-\left[ 8(e_{r0}+1)^{2}\sqrt{1-e_{r0}^{2}}\right. $ \\ 
& $\times \left( e_{r0}^{2}+2\right) e_{r0}^{3}\cos (\zeta _{1}-\zeta _{2})$
\\ 
& $+\cos \left( \zeta _{1}+\zeta _{2}\right) \left[
3176e_{r0}^{3}-3176e_{r0}^{2}\right.$ \\ 
& $+1552e_{r0}^{4}-1552e_{r0}^{5}-35e_{r0}^{6}$ \\ 
& $+\sqrt{1-e_{r0}^{2}}\left( 2376e_{r0}^{2}-2328e_{r0}^{3}\right.$ \\ 
& $-468e_{r0}^{4}+606e_{r0}^{5}-35e_{r0}^{6}$ \\ 
& $\left. \left. +92e_{r0}^{7}+1600e_{r0}-1600\right) \right]$ \\ 
\hline\hline
\end{tabular}%
\end{center}
\end{table}

\subsection{Initial shape variables in terms of the secular shape variables 
\label{inverting}}

The contributions to the averaged shape variables in terms of $\mathfrak{l}%
_{r0}$ and $e_{r0}$, Eqs. (\ref{lraverexpr}) and (\ref{eraverexpr}), were
presented in the previous subsection. Here, we invert these relations to
generate $\mathfrak{l}_{r0}$ and $e_{r0}$ in terms of $\mathfrak{\bar{l}}%
_{r} $\ and $\bar{e}_{r}$.

We do this in two steps. First, we take the perturbations to linear order.
Using Eqs. (\ref{lraverexpr}), (\ref{lravernewtexpr}), and (\ref{tau0expr})
for $\mathfrak{l}_{r0}$, we find 
\begin{eqnarray}
\mathfrak{l}_{r0} &=&\mathfrak{\bar{l}}_{r}\frac{\mathfrak{T}}{\mathfrak{T}%
_{0}}-\frac{\mathfrak{1}}{\mathfrak{T}_{0}}\left( \mathfrak{\bar{l}}_{rPN}+%
\mathfrak{\bar{l}}_{rSO}\right.  \notag \\
&&\left. +\mathfrak{\bar{l}}_{rSS}+\mathfrak{\bar{l}}_{rQM}\right) ~.
\end{eqnarray}%
Using Eqs (\ref{eraverexpr}), (\ref{eravernewtexpr}), and (\ref{tau0expr})
for $e_{r0}$, we get%
\begin{eqnarray}
e_{r0} &=&\bar{e}_{r}\frac{\mathfrak{T}}{\mathfrak{T}_{0}}-\frac{\mathfrak{1}%
}{\mathfrak{T}_{0}}\left( \bar{e}_{rPN}+\bar{e}_{rSO}\right.  \notag \\
&&\left. +\bar{e}_{rSS}+\bar{e}_{rQM}\right) ~.
\end{eqnarray}%
In the perturbation terms, we can insert the leading-order terms of $%
\mathfrak{l}_{r0}$ and $e_{r0}$, which are%
\begin{eqnarray}
\mathfrak{l}_{r0} &=&\mathfrak{\bar{l}}_{r}~,  \label{newtonianpartlr} \\
e_{r0} &=&\bar{e}_{r}~.  \label{newtonianparter}
\end{eqnarray}%
The results are%
\begin{equation}
\mathfrak{l}_{r0PN}=\frac{2\bar{e}_{r}(\bar{e}_{r}+1)(\eta -2)}{\mathfrak{%
\bar{l}}_{r}}~,  \label{lr0pn}
\end{equation}%
\begin{eqnarray}
\mathfrak{l}_{r0SO} &=&-\frac{\bar{e}_{r}\left( 1-\bar{e}_{r}^{2}\right)
\eta }{2(\bar{e}_{r}-1)\mathfrak{\bar{l}}_{r}^{2}}  \notag \\
&&\times \sum_{k=1}^{2}\left( 4^{2k-3}+3\right) \chi _{k}\cos \kappa _{k}~,
\end{eqnarray}%
\begin{eqnarray}
\mathfrak{l}_{r0SS} &=&-\frac{\chi _{1}\chi _{2}\eta }{32(1-\bar{e}_{r})\bar{%
e}_{r}^{2}\mathfrak{\bar{l}}_{r}^{3}}\left( L_{0}^{SS}\cos \kappa _{1}\cos
\kappa _{2}\right.  \notag \\
&&\left. -K_{0}^{SS}\sin \kappa _{1}\sin \kappa _{2}\right) ~,
\label{lr0SSb}
\end{eqnarray}%
\begin{eqnarray}
\mathfrak{l}_{r0QM} &=&\frac{\eta }{512\bar{e}_{r}\left( \bar{e}%
_{r}^{2}-1\right) \mathfrak{\bar{l}}_{r}^{3}}\sum_{k=1}^{2}\chi _{k}^{2}\nu
^{2k-3}w_{k}  \notag \\
&&\times \left[ -4\left( 47\bar{e}_{r}^{5}+1338\bar{e}_{r}^{4}+1446\bar{e}%
_{r}^{3}\right. \right.  \notag \\
&&\left. +3444\bar{e}_{r}^{2}+1264\bar{e}_{r}+960\right)  \notag \\
&&\times \sin ^{2}\kappa _{k}\cos 2\zeta _{k}+16\left( 5\bar{e}_{r}^{5}+33%
\bar{e}_{r}^{4}\right.  \notag \\
&&\left. +61\bar{e}_{r}^{3}+84\bar{e}_{r}^{2}+42\bar{e}_{r}+12\right)  \notag
\\
&&\left. \times \left( 3\cos 2\kappa _{k}+\allowbreak 1\right) \right] ~,
\label{lr0qm}
\end{eqnarray}%
where the coefficients $L_{0}^{SS}$ and $K_{0}^{SS}$ are listed in Table \ref%
{lr0kifcoeffs}. For $e_{r0}$, we have%
\begin{eqnarray}
e_{r0PN} &=&\frac{1}{\bar{e}_{r}\mathfrak{\bar{l}}_{r}^{2}}\left\{ -\left( 1-%
\bar{e}_{r}^{2}\right) ^{3/2}\left( 3\eta -5\right) \right.  \notag \\
&&+\frac{\left( 1+\bar{e}_{r}\right) ^{2}}{2}\left[ \bar{e}_{r}^{2}\left(
11\eta -14\right) \right.  \notag \\
&&\left. \left. -2\bar{e}_{r}\left( 5\eta -7\right) +6\eta -10\right]
\right\} ~,  \label{er0pn}
\end{eqnarray}%
\begin{eqnarray}
e_{r0SO} &=&-\frac{(\bar{e}_{r}+1)\left( 1-\bar{e}_{r}^{2}\right) \eta }{2%
\mathfrak{\bar{l}}_{r}^{3}}  \notag \\
&&\times \sum_{k=1}^{2}\left( 4^{2k-3}+3\right) \chi _{k}\cos \kappa _{k}~,
\end{eqnarray}%
\begin{eqnarray}
e_{r0SS} &=&-\frac{3\chi _{1}\chi _{2}\eta }{32(1-\bar{e}_{r})\bar{e}_{r}^{3}%
\mathfrak{\bar{l}}_{r}^{4}}\left( E_{0}^{SS}\cos \kappa _{1}\cos \kappa
_{2}\right.  \notag \\
&&\left. -F_{0}^{SS}\sin \kappa _{1}\sin \kappa _{2}\right) ~,
\label{er0SSb}
\end{eqnarray}%
\begin{eqnarray}
e_{r0QM} &=&\frac{\eta }{256\left( \bar{e}_{r}^{2}-1\right) \mathfrak{\bar{l}%
}_{r}^{4}}\sum_{k=1}^{2}\chi _{k}^{2}\nu ^{2k-3}w_{k}  \notag \\
&&\times \left[ -4\left( 10\bar{e}_{r}^{5}+753\bar{e}_{r}^{4}+712\bar{e}%
_{r}^{3}\right. \right.  \notag \\
&&\left. +1986\bar{e}_{r}^{2}+664\bar{e}_{r}+672\right)  \notag \\
&&\times \sin ^{2}\kappa _{k}\cos 2\zeta _{k}+8\left( 4\bar{e}_{r}^{5}+39%
\bar{e}_{r}^{4}\right.  \notag \\
&&+59\bar{e}_{r}^{3}+102\bar{e}_{r}^{2}+42\bar{e}_{r}  \notag \\
&&\left. \left. +24\right) \left( 3\cos 2\kappa _{k}+\allowbreak 1\right) 
\right] ~.  \label{er0qm}
\end{eqnarray}%
where the coefficients $E_{0}^{SS}$ and $F_{0}^{SS}$ can be found in Table %
\ref{er0kifcoeffs}. 
\begin{table}[tbp]
\caption{The coefficients in Eqs. (\protect\ref{lr0SSb}) and (\protect\ref%
{lr02pn}).}
\label{lr0kifcoeffs}
\begin{center}
\begin{tabular}{cc}
\hline\hline
$Coefficient$ & $Expression$ \\ \hline\hline
$L_{0,2}^{2PN}$ & $3\left( 75\eta ^{2}-176\eta +216\right)$ \\ \hline
$L_{0,1}^{2PN}$ & $48\left( 3\eta ^{2}-11\eta +10\right)$ \\ \hline
$L_{0,0}^{2PN}$ & $86\eta ^{2}-260\eta +176 $ \\ \hline
$K_{0,4}^{2PN}$ & $12\left( 12\eta ^{2}-39\eta +28\right)$ \\ \hline
$K_{0,3}^{2PN}$ & $-6\left( 18\eta ^{2}-63\eta +64\right)$ \\ \hline
$K_{0,2}^{2PN}$ & $3\left( 17\eta ^{2}-7\eta +28\right)$ \\ \hline
$K_{0,1}^{2PN}$ & $-2\left( 7\eta ^{2}+2\eta -32\right)$ \\ \hline
$K_{0,0}^{2PN}$ & $43\eta ^{2}-130\eta +88$ \\ \hline\hline
$L_{0}^{SS}$ & $4\bar{e}_{r}\left( 4\bar{e}_{r}^{4}+53\bar{e}_{r}^{3}\right.$
\\ 
& $\left. +78\bar{e}_{r}^{2}+72\bar{e}_{r}+24\right)$ \\ \hline
$K_{0}^{SS}$ & $\left[ -95\bar{e}_{r}^{5}+171\bar{e}_{r}^{4}-96\bar{e}%
_{r}^{3}\right.$ \\ 
& $+236\bar{e}_{r}^{2}-56\bar{e}_{r}+32$ \\ 
& $\left. -32\left( 1-\bar{e}_{r}\right) \left( 1-\bar{e}_{r}^{2}\right)
^{3/2}\right]$ \\ 
& $\times \cos (\zeta _{1}+\zeta _{2})+96\bar{e}_{r}^{2} $ \\ 
& $\times \left( 1+\bar{e}_{r}\right) ^{2}\left( 2\cos \zeta _{1}\cos \zeta
_{2}\right.$ \\ 
& $\left. -\sin \zeta _{1}\sin \zeta _{2}\right) +2\bar{e}_{r}\left( 4\bar{e}%
_{r}^{4}\right.$ \\ 
& $\left. +29\bar{e}_{r}^{3}+30\bar{e}_{r}^{2}+48\bar{e}_{r}+24\right) $ \\ 
& $\times \cos (\zeta _{1}-\zeta _{2})$ \\ \hline\hline
\end{tabular}%
\end{center}
\end{table}
\begin{table}[tbp]
\caption{The coefficients in Eqs. (\protect\ref{er0SSb}) and (\protect\ref%
{er02pn}).}
\label{er0kifcoeffs}
\begin{center}
\begin{tabular}{cc}
\hline\hline
$Coefficient$ & $Expression$ \\ \hline\hline
$E_{0,4}^{2PN}$ & $15\left( 2915\eta ^{2}-8904\eta +6192\right)$ \\ \hline
$E_{0,3}^{2PN}$ & $960\left( 36\eta ^{2}-102\eta +65\right)$ \\ \hline
$E_{0,2}^{2PN}$ & $4\left( 7673\eta ^{2}-20110\eta +15360\right)$ \\ \hline
$E_{0,1}^{2PN}$ & $960\left( 2\eta ^{2}-11\eta +15\right)$ \\ \hline
$E_{0,0}^{2PN}$ & $-8\left( 4559\eta ^{2}-13390\eta +9540\right)$ \\ \hline
$F_{0,6}^{2PN}$ & $15\left( 383\eta ^{2}-989\eta +560\right)$ \\ \hline
$F_{0,5}^{2PN}$ & $-30\left( 51\eta ^{2}-261\eta +224\right)$ \\ \hline
$F_{0,4}^{2PN}$ & $-15\left( 68\eta ^{2}-301\eta +232\right)$ \\ \hline
$F_{0,3}^{2PN}$ & $2\left( 893\eta ^{2}-3505\eta +3360\right)$ \\ \hline
$F_{0,2}^{2PN}$ & $-2\left( 1756\eta ^{2}-6425\eta +5250\right)$ \\ \hline
$F_{0,1}^{2PN}$ & $9358\eta ^{2}-28100\eta +20880$ \\ \hline
$F_{0,0}^{2PN}$ & $-4559\eta ^{2}+13390\eta -9540$ \\ \hline\hline
$E_{0}^{SS}$ & $16\bar{e}_{r}^{3}\left( 1+\bar{e}_{r}\right) \left( \bar{e}%
_{r}^{3}+3\bar{e}_{r}^{2}+4\bar{e}_{r}+2\right)$ \\ \hline
$F_{0}^{SS}$ & $\cos \left( \zeta _{1}+\zeta _{2}\right) \left[ 92\bar{e}%
_{r}^{7}-35\bar{e}_{r}^{6}\right.$ \\ 
& $+606\bar{e}_{r}^{5}-468\bar{e}_{r}^{4}-2328\bar{e}_{r}^{3}$ \\ 
& $+2376\bar{e}_{r}^{2}+1600\bar{e}_{r}-1600 $ \\ 
& $\left. +8\left( \bar{e}_{r}+1\right) \left( 3\bar{e}_{r}^{2}+200\right)
\left( 1-\bar{e}_{r}\right) ^{5/2}\right]$ \\ 
& $+32(\bar{e}_{r}+1)^{2}\bar{e}_{r}^{4} $ \\ 
& $\times \left( 2\cos \zeta _{1}\cos \zeta _{2}-\sin \zeta _{1}\sin \zeta
_{2}\right)$ \\ 
& $+8(\bar{e}_{r}+1)^{2}\left( \bar{e}_{r}^{2}+2\right) \bar{e}_{r}^{3} $ \\ 
& $\times \cos (\zeta _{1}-\zeta _{2})$ \\ \hline\hline
\end{tabular}%
\end{center}
\end{table}

The second step is to derive the 2PN terms. For this, we use Eqs. (\ref%
{lraverexpr}), (\ref{lravernewtexpr}), and (\ref{tau0expr}) for $\mathfrak{l}%
_{r0}$ and Eqs. (\ref{eraverexpr}), (\ref{eravernewtexpr}), and (\ref%
{tau0expr}) for $e_{r0}$: 
\begin{eqnarray}
\mathfrak{l}_{r0} &=&\mathfrak{\bar{l}}_{r}\frac{\mathfrak{T}}{\mathfrak{T}%
_{0}}-\frac{\mathfrak{1}}{\mathfrak{T}_{0}}\left( \mathfrak{\bar{l}}_{rPN}+%
\mathfrak{\bar{l}}_{r2PN}\right) ~, \\
e_{r0} &=&\bar{e}_{r}\frac{\mathfrak{T}}{\mathfrak{T}_{0}}-\frac{\mathfrak{1}%
}{\mathfrak{T}_{0}}\left( \bar{e}_{rPN}+\bar{e}_{r2PN}\right) ~.
\end{eqnarray}%
This time, in order to get the 2PN terms, we need the previously calculated
1PN expressions of $\mathfrak{l}_{r0}$ and $e_{r0}$. After the Taylor
expansion to 2PN order, we find%
\begin{eqnarray}
\mathfrak{l}_{r02PN} &=&-\frac{(1-\bar{e}_{r}^{2})^{3/2}}{24\bar{e}_{r}^{2}%
\mathfrak{\bar{l}}_{r}^{3}}\sum_{k=0}^{2}L_{0,k}^{2PN}\bar{e}_{r}^{k}  \notag
\\
&&+\frac{(\bar{e}_{r}+1)^{2}}{12\bar{e}_{r}^{2}\mathfrak{\bar{l}}_{r}^{3}}%
\sum_{k=0}^{4}K_{0,k}^{2PN}\bar{e}_{r}^{k}~,  \label{lr02pn}
\end{eqnarray}%
\begin{eqnarray}
e_{r02PN} &=&-\frac{\left( 1-\bar{e}_{r}^{2}\right) ^{3/2}}{960\bar{e}%
_{r}^{3}\mathfrak{\bar{l}}_{r}^{4}}\sum_{k=0}^{4}E_{0,k}^{2PN}\bar{e}_{r}^{k}
\notag \\
&&+\frac{(1+\bar{e}_{r})^{2}}{120\bar{e}_{r}^{3}\mathfrak{\bar{l}}_{r}^{4}}%
\sum_{k=0}^{6}F_{0,k}^{2PN}\bar{e}_{r}^{k}~.  \label{er02pn}
\end{eqnarray}%
The coefficients $L_{0,k}^{2PN}$ and $K_{0,k}^{2PN}$ are given in Table \ref%
{lr0kifcoeffs}, while the terms $E_{0,k}^{2PN}$ and $F_{0,k}^{2PN}$ can be
found in Table \ref{er0kifcoeffs}.

The full expressions of $\mathfrak{l}_{r0}$ and $e_{r0}$ are the sum of the
corresponding above contributions Eqs. (\ref{newtonianpartlr}), (\ref{lr0pn}%
)--(\ref{lr0qm}), (\ref{lr02pn}) and Eqs. (\ref{newtonianparter}), (\ref%
{er0pn})--(\ref{er0qm}), (\ref{er02pn}), respectively.

\subsection{Radial period in terms of time averages of shape variables \label%
{timeeq}}

Here, we finally are able to express the radial period in terms of averaged
quantities by replacing the initial values at the periastron with time
averages over $\chi _{p}\in \left[ 0,2\pi \right] $. The various order
contributions to Eq. (\ref{periodaver}) become

\begin{eqnarray}
\mathfrak{\tilde{T}} &=&\frac{2\mathfrak{\bar{l}}_{r}^{3}\pi }{\left( 1-\bar{%
e}_{r}^{2}\right) ^{3/2}}~, \\
\tilde{\tau}_{PN} &=&\sqrt{1-\bar{e}_{r}^{2}}(15-9\eta )  \notag \\
&&+\left( 1-\bar{e}_{r}^{2}\right) (7\eta -6)~, \\
\tilde{\tau}_{SO} &=&0~, \\
\tilde{\tau}_{2PN} &=&\frac{\sqrt{1-\bar{e}_{r}^{2}}}{64\bar{e}_{r}^{4}}%
\sum_{k=0}^{6}\bar{U}_{k}\bar{e}_{r}^{k}  \notag \\
&&-\frac{(\bar{e}_{r}+1)}{8\bar{e}_{r}^{4}}\sum_{k=0}^{7}\bar{V}_{k}\bar{e}%
_{r}^{k}~,  \label{tau2pnkif} \\
\tilde{\tau}_{QM} &=&\frac{3\eta }{512\bar{e}_{r}\left( 1-\bar{e}%
_{r}^{2}\right) ^{2}}\sum_{k=1}^{2}\chi _{k}^{2}\nu ^{2k-3}w_{k}  \notag \\
&&\times \left[ \bar{U}^{QM}\sin ^{2}\kappa _{k}\cos 2\zeta _{k}\right. 
\notag \\
&&\left. +\bar{V}^{QM}\left( 3\cos 2\kappa _{k}\allowbreak +1\right) \right]
~, \\
\bar{\tau}_{SS} &=&\frac{3\chi _{1}\chi _{2}\eta }{8(1-\bar{e}_{r})^{2}\bar{e%
}_{r}}\left( \cos \kappa _{1}\cos \kappa _{2}\bar{U}^{SS}\right.  \notag \\
&&\left. +\sin \kappa _{1}\sin \kappa _{2}\bar{V}^{SS}\right) ~.
\label{tausskif}
\end{eqnarray}%
$\allowbreak $The coefficients in the above expressions are listed in Table %
\ref{tauaver2pn}. 
\begin{table}[tbp]
\caption{The coefficients in Eqs. (\protect\ref{tau2pnkif})--(\protect\ref%
{tausskif}).}
\label{tauaver2pn}
\begin{center}
\begin{tabular}{cc}
\hline\hline
$Coefficient$ & $Expression$ \\ \hline\hline
$\bar{U}_{6}$ & $-437\eta ^{2}+3336\eta -1008$ \\ \hline
$\bar{U}_{5}$ & $-64\left( 8\eta ^{2}-6\eta -5\right) $ \\ \hline
$\bar{U}_{4}$ & $-8\left( 211\eta ^{2}-159\eta +336\right) $ \\ \hline
$\bar{U}_{3}$ & $64\left( 4\eta ^{2}+11\eta -5\right)$ \\ \hline
$\bar{U}_{2}$ & $-8\left( 79\eta ^{2}-600\eta +528\right)$ \\ \hline
$\bar{U}_{1}$ & $-128\left( \eta ^{2}-8\eta +15\right)$ \\ \hline
$\bar{U}_{0}$ & $32\left( 65\eta ^{2}-238\eta +180\right)$ \\ \hline\hline
$\bar{V}_{7}$ & $224\eta ^{2}-690\eta +360$ \\ \hline
$\bar{V}_{6}$ & $2\left( 64\eta ^{2}-11\eta -12\right) $ \\ \hline
$\bar{V}_{5}$ & $139\eta ^{2}-410\eta +452$ \\ \hline
$\bar{V}_{4}$ & $-179\eta ^{2}+266\eta -308$ \\ \hline
$\bar{V}_{3}$ & $-27\eta ^{2}+28\eta +8$ \\ \hline
$\bar{V}_{2}$ & $67\eta ^{2}-4\eta +72 $ \\ \hline
$\bar{V}_{1}$ & $-12\left( 23\eta ^{2}-90\eta +80\right) $ \\ \hline
$\bar{V}_{0}$ & $260\eta ^{2}-952\eta +720$ \\ \hline\hline
$\bar{U}^{QM}$ & $-4\left( 27\bar{e}_{r}^{7}-72\bar{e}_{r}^{6}+263\bar{e}%
_{r}^{5}\right.$ \\ 
& $-1674\bar{e}_{r}^{4}-1702\bar{e}_{r}^{3}-4116\bar{e}_{r}^{2}$ \\ 
& $\left. -1360\bar{e}_{r}-960\right)$ \\ \hline
$\bar{V}^{QM}$ & $16\left( \bar{e}_{r}^{7}-2\bar{e}_{r}^{6}+9\bar{e}%
_{r}^{5}-43\bar{e}_{r}^{4}\right.$ \\ 
& $\left. -69\bar{e}_{r}^{3}-108\bar{e}_{r}^{2}-46\bar{e}_{r}-12\right)$ \\ 
\hline\hline
$\bar{U}^{SS}$ & $-8\bar{e}_{r}^{5}+21\bar{e}_{r}^{4}-15\bar{e}_{r}^{3}$ \\ 
& $-38\bar{e}_{r}^{2}-56\bar{e}_{r}-24$ \\ \hline
$\bar{V}^{SS}$ & $\frac{\cos (\zeta _{1}-\zeta _{2})}{2}\allowbreak \left( 8%
\bar{e}_{r}^{5}-21\bar{e}_{r}^{4}\right.$ \\ 
& $\left. +15\bar{e}_{r}^{3}+38\bar{e}_{r}^{2}+56\bar{e}_{r}+24\right)$ \\ 
& $+\frac{\cos (\zeta _{1}+\zeta _{2})}{4\bar{e}_{r}^{2}(\bar{e}_{r}+1)}%
\left[ 371\bar{e}_{r}^{7}-276\bar{e}_{r}^{6}\right.$ \\ 
& $+\bar{e}_{r}^{5}\left( -48\bar{e}_{r}+104\sqrt{1-\bar{e}_{r}^{2}}%
+1771\right)$ \\ 
& $-\bar{e}_{r}^{4}\left( 48\bar{e}_{r}+104\sqrt{1-\bar{e}_{r}^{2}}%
+1517\right)$ \\ 
& $+8\bar{e}_{r}^{3}\left( 24\bar{e}_{r}+583\sqrt{1-\bar{e}_{r}^{2}}%
-854\right)$ \\ 
& $+4\bar{e}_{r}^{2}\left( 48\bar{e}_{r}-1166\sqrt{-\bar{e}_{r}^{2}+1}%
+1881\right)$ \\ 
& $-8\left( 596\sqrt{1-\bar{e}_{r}^{2}}-593\right) \bar{e}_{r}$ \\ 
& $\left. +4768\left( \sqrt{1-\bar{e}_{r}^{2}}-1\right) \right]$ \\ 
\hline\hline
\end{tabular}%
\end{center}
\end{table}

\subsection{Expansion of the averaged PN parameter}

The PN parameter associated to the averaged dynamics is given by%
\begin{equation}
\bar{\varepsilon}=\bar{\varepsilon}_{0PN}+\bar{\varepsilon}_{1PN}+\bar{%
\varepsilon}_{2PN}+\bar{\varepsilon}_{SO}+\bar{\varepsilon}_{SS}+\bar{%
\varepsilon}_{QM}~,  \label{epsaver}
\end{equation}%
with

\begin{equation}
\bar{\varepsilon}_{0PN}=\frac{2\mathfrak{\bar{l}}_{r}\pi }{\mathfrak{T}%
\left( 1-\bar{e}_{r}^{2}\right) ^{1/2}}~,
\end{equation}%
\begin{equation}
\bar{\varepsilon}_{1PN}=-\frac{\pi }{\mathfrak{T\bar{l}}_{r}}\left[ \left(
6-7\eta \right) \left( 1-\bar{e}_{r}^{2}\right) ^{1/2}-9\left( 2-\eta
\right) \right] ~,
\end{equation}%
\begin{eqnarray}
\bar{\varepsilon}_{SO} &=&\frac{3\pi }{2\mathfrak{T\bar{l}}_{r}^{2}}\left( 1-%
\frac{2}{3}\sqrt{\frac{1+\bar{e}_{r}}{1-\bar{e}_{r}}}\right)  \notag \\
&&\times \sum_{i=1}^{2}\chi _{i}\cos \kappa _{i}\left( 4\nu ^{2i-3}+3\right)
~,
\end{eqnarray}%
\begin{eqnarray}
\bar{\varepsilon}_{SS} &=&\frac{\pi \chi _{1}\chi _{2}\eta }{16\bar{e}%
_{r}^{2}(\bar{e}_{r}-1)^{3}(\bar{e}_{r}+1)^{2}\mathfrak{\bar{l}}_{r}^{3}%
\mathfrak{T}}  \notag \\
&&\times \left\{ 16\bar{e}_{r}\left( \bar{e}_{r}+1\right) \cos \kappa
_{1}\cos \kappa _{2}M^{SS}\right.  \notag \\
&&\left. +\sin (\kappa _{1})\sin (\kappa _{2})N^{SS}\right\}
\label{epsaverSS}
\end{eqnarray}%
\begin{eqnarray}
\bar{\varepsilon}_{QM} &=&\frac{\pi \eta }{256\bar{e}_{r}\left( 1-\bar{e}%
_{r}^{2}\right) ^{5/2}\mathfrak{\bar{l}}_{r}^{3}\mathfrak{T}}%
\sum_{i=1}^{2}\nu ^{2i-3}\chi _{i}^{2}w_{i}  \notag \\
&&\times \left\{ M^{QM}\left[ \cos \left( 2\kappa _{i}-2\zeta _{i}\right)
+\cos \left( 2\kappa _{i}+2\zeta _{i}\right) \right. \right.  \notag \\
&&\left. \left. -2\cos 2\zeta _{i}\right] +N^{QM}\right\}  \label{epsaverQM}
\end{eqnarray}%
The coefficients of $\bar{\varepsilon}_{SS}$ and $\bar{\varepsilon}_{QM}$
are given in Table \ref{espcoeffs1}. 
\begin{table}[tbp]
\caption{The coefficients $M^{SS}$,$~M^{QM}$,$~N^{SS}$, and $N^{QM}$ of $%
\mathfrak{\bar{\protect\varepsilon}}$ in Eqs. (\protect\ref{epsaverSS}) and (%
\protect\ref{epsaverQM}).}
\label{espcoeffs1}
\begin{center}
\begin{tabular}{cc}
\hline\hline
$Coefficient$ & $Expression$ \\ \hline\hline
$M^{SS}$ & $-2\bar{e}_{r}+4\bar{e}_{r}^{2}-4\bar{e}_{r}^{4}+2\bar{e}_{r}^{5}$
\\ 
& $-\left( 16\bar{e}_{r}^{2}+190+96-19\bar{e}_{r}^{5}-203\bar{e}%
_{r}^{4}\right. $ \\ 
& $\left. -110\bar{e}_{r}^{3}\right) \sqrt{1-\bar{e}_{r}^{2}}$ \\ \hline
$N^{SS}$ & $\cos \left( \zeta _{1}+\zeta _{2}\right) \left\{ \allowbreak
512+\allowbreak 2048\bar{e}_{r}^{2}\right. $ \\ 
& $-\allowbreak 3072\bar{e}_{r}^{4}+2048\bar{e}_{r}^{6}-512\bar{e}_{r}^{8}$
\\ 
& $+\left[ 5312-\allowbreak 5696\bar{e}_{r}-\allowbreak 1608\bar{e}%
_{r}^{2}\right. $ \\ 
& $+\allowbreak 6248\bar{e}_{r}^{3}+\allowbreak 220\bar{e}%
_{r}^{4}\allowbreak -1994\bar{e}_{r}^{5}$ \\ 
& $-5031\bar{e}_{r}^{6}+1244\bar{e}_{r}^{7}$ \\ 
& $+\left( -4800+\allowbreak 4728\bar{e}_{r}^{2}+72\bar{e}_{r}^{4}\right) $
\\ 
& $\left. \left. \times \left( 1-\bar{e}_{r}\right) ^{3/2}\right] \sqrt{1-%
\bar{e}_{r}^{2}}\right\} $ \\ 
& $+8\bar{e}_{r}(\bar{e}_{r}+1)\cos \left( \zeta _{1}-\zeta _{2}\right) $ \\ 
& $\left[ -2\bar{e}_{r}\allowbreak -4\bar{e}_{r}^{2}-4\bar{e}_{r}^{4}+2\bar{e%
}_{r}^{5}\right. $ \\ 
& $\left( -96-190\bar{e}_{r}-176\bar{e}_{r}^{2}-82\bar{e}_{r}^{3}\right. $
\\ 
& $\left. \left. +203\bar{e}_{r}^{4}+19\bar{e}_{r}^{5}\right) \sqrt{1-\bar{e}%
_{r}^{2}}\right] $ \\ \hline\hline
$M^{QM}$ & $27\bar{e}_{r}^{7}-40\bar{e}_{r}^{6}+167\bar{e}_{r}^{5}-2122\bar{e%
}_{r}^{4}$ \\ 
& $-1990\bar{e}_{r}^{3}-3700\bar{e}_{r}^{2}-976\bar{e}_{r}-960$ \\ \hline
$N^{QM}$ & $16\bar{e}_{r}^{7}+416\bar{e}_{r}^{6}+1104\bar{e}_{r}^{5}-816\bar{%
e}_{r}^{4}$ \\ 
& $-2064\bar{e}_{r}^{3}-2048\bar{e}_{r}^{2}-736\bar{e}_{r}-192$ \\ 
& $+\left( 64\bar{e}_{r}+64\bar{e}_{r}^{5}-128\bar{e}_{r}^{3}\right) \sqrt{1-%
\bar{e}_{r}^{2}}$ \\ 
& $-256\left( 7\bar{e}_{r}^{4}+15\bar{e}_{r}^{3}-2\bar{e}_{r}^{2}\right.$ \\ 
& $\left. -15\bar{e}_{r}-5\right) \bar{e}_{r}^{2}\sin ^{2}\kappa _{i}\cos
^{2}\zeta _{i}$ \\ 
& $+16\left\{ 3\bar{e}_{r}^{7}-34\bar{e}_{r}^{6}-121\bar{e}_{r}^{4}-304\bar{e%
}_{r}^{2}\right.$ \\ 
& $-138\bar{e}_{r}-33\bar{e}_{r}^{5}-\allowbreak 147\bar{e}_{r}^{3}-36$ \\ 
& $\left. +\left( -24+12\bar{e}_{r}+12\bar{e}_{r}^{5}\right) \sqrt{1-\bar{e}%
_{r}^{2}}\right] \cos 2\kappa _{i}$ \\ \hline\hline
\end{tabular}%
\end{center}
\end{table}
\begin{eqnarray}
\bar{\varepsilon}_{2PN} &=&\frac{-\pi }{4915200\left( 1-\bar{e}%
_{r}^{2}\right) ^{5/2}\bar{e}_{r}^{4}\mathfrak{\bar{l}}^{3}\mathfrak{T}} 
\notag \\
&&\times \left\{ 15\bar{e}_{r}^{5}M_{1}^{2PN}-2457600\left( 1+\bar{e}%
_{r}\right) ^{3}\bar{e}_{r}^{3}M_{2}^{2PN}\right.  \notag \\
&&-10240\bar{e}_{r}^{2}\left( 1+\bar{e}_{r}\right) ^{2}M_{3}^{2PN}  \notag \\
&&+2\left( 1+\bar{e}_{r}\right) \left[ 8\left( 1-\bar{e}_{r}-\bar{e}_{r}^{2}+%
\bar{e}_{r}^{3}\right) \sqrt{1-\bar{e}_{r}^{2}}\right.  \notag \\
&&\left. -8+8\bar{e}_{r}+12\bar{e}_{r}^{2}-14\bar{e}_{r}^{3}-\bar{e}_{r}^{4}%
\right] M_{4}^{2PN}  \notag \\
&&+5\left[ 15\bar{e}_{r}^{4}-20\bar{e}_{r}^{2}-8\left( 1-\bar{e}%
_{r}^{2}\right) ^{5/2}\right.  \notag \\
&&\left. \left. +8\right] M_{5}^{2PN}\right\}  \label{epsaver2PN}
\end{eqnarray}%
The coefficents of $\bar{\varepsilon}_{2PN}$ are enlisted in Table \ref%
{espcoeffs2}. 
\begin{table}[th]
\caption{The coefficients $M_{k}^{2PN}$ of $\mathfrak{\bar{\protect%
\varepsilon}}$ in Eq. (\protect\ref{epsaver2PN}).}
\label{espcoeffs2}
\begin{center}
\begin{tabular}{cc}
\hline\hline
$Coefficient$ & $Expression$ \\ \hline\hline
$M_{1}^{2PN}$ & $-15\eta (98967\eta +118819)\bar{e}_{r}^{6}$ \\ 
& $+1920(\eta (8399\eta -17134)+12016)\bar{e}_{r}^{5}$ \\ 
& $+4(\eta (5837763\eta -15119020)+8076560)\bar{e}_{r}^{4}$ \\ 
& $-5120(\eta (2431\eta -13254)+662)\bar{e}_{r}^{3}$ \\ 
& $+160(41\eta (3071\eta -11629)+502740)\bar{e}_{r}^{2}$ \\ 
& $+40960(\eta (68\eta +1067)-408)\bar{e}_{r}$ \\ 
& $+1482240(\eta -3)^{2}$ \\ \hline
$M_{2}^{2PN}$ & $8\left( 4\eta ^{2}-11\eta +6\right) \bar{e}_{r}^{5}$ \\ 
& $+\left( 161\eta ^{2}-394\eta +256\right) \bar{e}_{r}^{4}$ \\ 
& $+\left( 49\eta ^{2}-166\eta +148\right) \bar{e}_{r}^{3}$ \\ 
& $+2\left( -25\eta ^{2}+59\eta -48\right) \bar{e}_{r}^{2}$ \\ 
& $-2\left( -33\eta ^{2}+83\eta -52\right) \bar{e}_{r}$ \\ 
& $+4\left( 9\eta ^{2}-36\eta +35\right) $ \\ 
& $+\sqrt{1-\bar{e}_{r}^{2}}\left[ \left( -12\eta ^{2}+56\eta -64\right) 
\bar{e}_{r}^{4}\right. $ \\ 
& $+\left( 24\eta ^{2}-112\eta +128\right) \bar{e}_{r}^{3}$ \\ 
& $+2\left( 39\eta ^{2}-95\eta +48\right) \bar{e}_{r}^{2}$ \\ 
& $-2\left( 27\eta ^{2}-51\eta +10\right) \bar{e}_{r}$ \\ 
& $\left. -4\left( 9\eta ^{2}-36\eta +35\right) \right] $ \\ \hline
$M_{3}^{2PN}$ & $120\left( 133\eta ^{2}-349\eta +168\right) \bar{e}_{r}^{8}$
\\ 
& $-240\left( 33\eta ^{2}-155\eta +96\right) \bar{e}_{r}^{7}$ \\ 
& $+15\left( -2304\eta ^{2}+5856\eta -\allowbreak 3680\right) \bar{e}%
_{r}^{6} $ \\ 
& $-2\left( 17\,256\eta ^{2}+37\,520\eta +27\,520\right) \bar{e}_{r}^{5}$ \\ 
& $+\left( -14\,856\eta ^{2}+73\,720\eta -\allowbreak 47\,360\right) \bar{e}%
_{r}^{4}$ \\ 
& $-8\left( -8312\eta ^{2}+24050\eta -18\,080\right) \bar{e}_{r}^{3}$ \\ 
& $+4\left( -6554\eta ^{2}+12260\eta -\allowbreak 8720\right) \bar{e}%
_{r}^{2} $ \\ 
& $-16\left( 3709\eta ^{2}-11310\eta +8660\right) \bar{e}_{r}$ \\ 
& $+8\left( 3049\eta ^{2}-8490\eta +5660\right)$ \\ 
& $+\sqrt{1-\bar{e}_{r}^{2}}\left[15\left( 587\eta ^{2}- 2696\eta
+1264\right) \right.$ \\ 
& $-2\left( 8805\eta ^{2}- 40\,440\eta + 21\,360\right) \bar{e}_{r}^{5}$ \\ 
& $+\left( 6177\eta ^{2}-17\,840+ 27\,280\right) \bar{e}_{r}^{4}$ \\ 
& $-8\left( 5583\eta ^{2}-12110\eta +12\,280\right) \bar{e}_{r}^{3}$ \\ 
& $+4\left( 3085\eta ^{2}-1610\eta +360\right) \bar{e}_{r}^{2}$ \\ 
& $+16\left( 3709\eta ^{2}-11310\eta +8660\right) \bar{e}_{r}$ \\ 
& $\left.-8\left( 3049\eta ^{2}-8490\eta +5660\right) \right]$ \\ \hline
$M_{4}^{2PN}$ & $1200\eta (851\eta -3617)\bar{e}_{r}^{6}$ \\ 
& $+15\left( 1731801\eta ^{2}-3339132\eta +1884800\right) \bar{e}_{r}^{5}$
\\ 
& $+800\left( 37875\eta ^{2}-119180\eta +33904\right) \bar{e}_{r}^{4}$ \\ 
& $-96\left( 460104\eta ^{2}-1355255\eta +1016100\right) \bar{e}_{r}^{3}$ \\ 
& $-25600\left( 1177\eta ^{2}+5202\eta -2962\right) \bar{e}_{r}^{2}$ \\ 
& $-9600\left( 399\eta ^{2}-1814\eta +1851\right) \bar{e}_{r}$ \\ 
& $-153600\left( 43\eta ^{2}-174\eta +135\right) $ \\ \hline
$M_{5}^{2PN}$ & $15\eta (98967\eta +118819)\bar{e}_{r}^{6}$ \\ 
& $-1920\left( 8399\eta ^{2}-17134\eta +12016\right) \bar{e}_{r}^{5}$ \\ 
& $-4\left( 5837763\eta ^{2}-15119020\eta +8076560\right) \bar{e}_{r}^{4}$
\\ 
& $+5120\left( 2431\eta ^{2}-13254\eta +662\right) \bar{e}_{r}^{3}$ \\ 
& $-160\left( 125911\eta ^{2}-476789\eta +502740\right) \bar{e}_{r}^{2}$ \\ 
& $-40960\left( 68\eta ^{2}+1067\eta -408\right) \bar{e}_{r}$ \\ 
& $-1482240(\eta -3)^{2}$ \\ \hline\hline
\end{tabular}%
\end{center}
\end{table}

\section{Regular evolution despite a jump in $\Delta \protect\zeta $ when
one of the spins crosses the orbital angular momentum\label{reg}}

We note that when either of the spins becomes perpendicular to the orbit Eq.
(\ref{eq3}) blows up due to the $\sin ^{-1}\kappa _{i}$ and $\cot \kappa
_{i} $ factors, the angle $\Delta \zeta $ becoming ill defined. We show in
this subsection that this is but a coordinate singularity.

To illustrate this, we assume that $\sin \kappa _{2}\ll 1$, i.e.%
\begin{equation}
\kappa _{2}\left( \mathfrak{t}\right) =\kappa _{\left( 0\right) 2}+\delta
\kappa _{2}\left( \mathfrak{t}\right) ~,
\end{equation}%
with $\kappa _{\left( 0\right) 2}\in \left\{ 0,\pi \right\} $ and $%
\left\vert \delta \kappa _{2}\right\vert \ll 1$. As said before, we are
interested in the evolution of $\mathbf{S}_{\mathbf{2}}$ across the orbital
angular momentum direction. During this $\kappa _{1}$ behaves as a
quasiconstant (since $\dot{\kappa}_{1}\propto \sin \kappa _{2}$):%
\begin{equation}
\kappa _{1}\left( \mathfrak{t}\right) =\kappa _{\left( 0\right) 1}+\delta
\kappa _{1}\left( \mathfrak{t}\right) ~,
\end{equation}%
with $\kappa _{\left( 0\right) 1}$ a constant value and $\delta \kappa
_{1}\ll 1$. In addition we assume that $\mathbf{S}_{\mathbf{1}}$ points away
from $\mathbf{L}_{\mathbf{N}}$; thus, $\sin \kappa _{\left( 0\right)
1}>>\cos \kappa _{\left( 0\right) 1}\delta \kappa _{1}$. Then, the evolution
equations for $\Delta \zeta $, $\delta \kappa _{1}$, and $\delta \kappa _{2}$
to leading order become%
\begin{equation}
\overline{\frac{d\Delta \zeta }{d\mathfrak{t}}}=A_{\Delta \zeta }-B\frac{%
\cos \Delta \zeta }{\delta \kappa _{2}}~,  \label{DzetaEq1}
\end{equation}%
\begin{equation}
\overline{\frac{d\delta \kappa _{1}}{d\mathfrak{t}}}=\epsilon A_{\delta
\kappa _{1}}\delta \kappa _{2}\sin \Delta \zeta ~,  \label{dkappa1}
\end{equation}%
\begin{equation}
\overline{\frac{d\delta \kappa _{2}}{d\mathfrak{t}}}=-B\sin \Delta \zeta ~,
\label{dkappa2}
\end{equation}%
with $\epsilon =\cos \kappa _{\left( 0\right) 2}=\pm 1$ and the coefficients%
\begin{eqnarray}
\frac{A_{\Delta \zeta }}{R} &=&\nu -\frac{1}{\nu }-\epsilon \left( 2\nu
+1\right) x_{2}  \notag \\
&&+\left( \epsilon +\nu x_{2}\right) w_{2}x_{2}+\left( \frac{2}{\nu }%
+1-w_{1}\right.  \notag \\
&&\left. -\frac{w_{1}x_{1}}{\nu }\cos \kappa _{\left( 0\right) 1}\right)
x_{1}\cos \kappa _{\left( 0\right) 1}~,
\end{eqnarray}%
\begin{eqnarray}
\frac{B}{R} &=&\left( 1+\frac{1}{\nu }-\epsilon x_{2}\right.  \notag \\
&&\left. -\frac{w_{1}x_{1}}{\nu }\cos \kappa _{\left( 0\right) 1}\right)
x_{1}\sin \kappa _{\left( 0\right) 1}~,
\end{eqnarray}%
\begin{eqnarray}
\frac{A_{\delta \kappa _{1}}}{R} &=&\left( 1+\nu -\epsilon \nu
w_{2}x_{2}\right.  \notag \\
&&\left. -x_{1}\cos \kappa _{\left( 0\right) 1}\right) x_{2}~.
\end{eqnarray}%
Note that for notational simplicity we omitted the overbar form the secular
time derivatives.

From (\ref{DzetaEq1}) and (\ref{dkappa2}), we find 
\begin{equation}
\frac{d^{2}}{d\mathfrak{t}^{2}}\left[ \delta \kappa _{2}\sin \Delta \zeta %
\right] +A_{\Delta \zeta }^{2}\delta \kappa _{2}\sin \Delta \zeta =0~,
\end{equation}%
which gives%
\begin{equation}
\delta \kappa _{2}\sin \Delta \zeta =Q_{1}\cos \left( A_{\Delta \zeta }%
\mathfrak{t}+G\right) ~,  \label{sinDzeta}
\end{equation}%
with constants $Q_{1}>0$ and $G$. Then, Eq. (\ref{dkappa2}) results in%
\begin{equation}
\left( \delta \kappa _{2}\right) ^{2}=-\frac{2BQ_{1}}{A_{\Delta \zeta }}\sin
\left( A_{\Delta \zeta }\mathfrak{t}+G\right) +Q_{2}^{2}~,  \label{dkappa2sq}
\end{equation}%
with an integration constant $Q_{2}^{2}$. By substituting the solutions (\ref%
{sinDzeta}) and (\ref{dkappa2sq}) into Eqs. (\ref{DzetaEq1}) and (\ref%
{dkappa2}), we find the following relation: 
\begin{equation}
Q_{2}^{2}=Q_{1}^{2}+\frac{B^{2}}{A_{\Delta \zeta }^{2}}~.
\end{equation}%
With the notation%
\begin{equation}
C_{1}=\frac{A_{\Delta \zeta }}{B}Q_{1}~,
\end{equation}%
Eqs. (\ref{sinDzeta}) and (\ref{dkappa2sq}) become%
\begin{equation}
\left( \delta \kappa _{2}\right) ^{2}=\frac{B^{2}}{A_{\Delta \zeta }^{2}}%
\left[ 1+C_{1}^{2}-2C_{1}\sin \left( A_{\Delta \zeta }\mathfrak{t}+G\right) %
\right] ~,  \label{dkappa2sq2}
\end{equation}%
\begin{equation}
\sin \Delta \zeta =\frac{BC_{1}}{A_{\Delta \zeta }}\frac{\cos \left(
A_{\Delta \zeta }\mathfrak{\bar{t}}+G\right) }{\delta \kappa _{2}}~,
\label{sinDzetasmallk}
\end{equation}

The minimum value of $\left( \delta \kappa _{2}\right) ^{2}$ is given by $%
B^{2}\left( 1-C_{1}\right) ^{2}/A_{\Delta \zeta }^{2}$ and when this does
not vanish, we find from Eq. (\ref{sinDzetasmallk}) the following
restriction for the integration constant $C_{1}$:%
\begin{equation}
\left( \frac{C_{1}}{C_{1}-1}\right) ^{2}<1~.
\end{equation}

In the other case when $\delta \kappa _{2}$ can vanish, the expression (\ref%
{sinDzetasmallk}) is regular for $\delta \kappa _{2}\rightarrow 0$ only if $%
\cos \left( A_{\Delta \zeta }\mathfrak{\bar{t}}+G\right) \rightarrow 0$ at
the same time. From Eq. (\ref{dkappa2sq2}), we find that these conditions
can be satisfied only if $\sin \left( A_{\Delta \zeta }\mathfrak{t}+G\right)
\rightarrow 1$ for $\delta \kappa _{2}\rightarrow 0$ and the integration
constant $C_{1}$ is $1$. Then, the solutions read as%
\begin{equation}
\left( \delta \kappa _{2}\right) ^{2}=\frac{2B^{2}}{A_{\Delta \zeta }^{2}}%
\left[ 1-\sin \left( A_{\Delta \zeta }\mathfrak{t}+G\right) \right] ~,
\end{equation}%
\begin{equation}
\sin \Delta \zeta =\frac{B}{A_{\Delta \zeta }}\frac{\cos \left( A_{\Delta
\zeta }\mathfrak{t}+G\right) }{\delta \kappa _{2}}~.  \label{sinDzetaReg}
\end{equation}%
With this expression, from (\ref{dkappa1}), we have%
\begin{equation}
\delta \kappa _{1}=\epsilon \frac{A_{\delta \kappa _{1}}B}{A_{\Delta \zeta
}^{2}}\sin \left( A_{\Delta \zeta }\mathfrak{t}+G\right) +D~,
\end{equation}%
with $D$ an integration constant.

For $A_{\Delta \zeta }\mathfrak{t}+G\rightarrow M\pi /2$, where the integer $%
M$ is chosen such that $\delta \kappa _{2}\rightarrow 0$ at the same time,
we find%
\begin{equation}
\left( \delta \kappa _{2}\right) ^{2}=\frac{B^{2}}{A_{\Delta \zeta }^{2}}%
y^{2}\left( 1-\frac{y^{2}}{12}+\mathcal{O}\left( y^{4}\right) \right) ~,
\label{dkappa2sqsmall}
\end{equation}%
\begin{equation}
\sin ^{2}\Delta \zeta =1-\frac{y^{2}}{4}+\mathcal{O}\left( y^{4}\right) ~,
\label{sinsqDzetasmall}
\end{equation}%
with%
\begin{equation}
y=A_{\Delta \zeta }\mathfrak{t}+G-M\frac{\pi }{2}~.
\end{equation}%
Equations (\ref{dkappa2sqsmall}) and (\ref{sinsqDzetasmall}) show that, with
the exception of the case when $\kappa _{1}$ vanishes but otherwise for
general configurations, $\Delta \zeta \rightarrow \pm \pi /2$ as$~\kappa
_{2}\rightarrow \left\{ 0,\pi \right\} $. According to the definition of
polar spin angle, $\delta \kappa _{2}$ does not change sign when the spin
crosses the axis. Thus, from Eq. (\ref{sinDzetaReg}), we find that $\sin
\Delta \zeta $ must change sign as $\delta \kappa _{2}$ vanishes, implying a
jump of $\Delta \zeta $ by $\pi $ whenever $\mathbf{S}_{2}$ goes through the
axis defined by $\mathbf{L}_{\mathbf{N}}$.

Thus, we have proven that both $\cos \Delta \zeta \propto y$ and $\delta
\kappa _{2}\propto y$; thus, those terms in Eq. (\ref{eq3}) which contain a
factor of $\cos \Delta \zeta /\sin \kappa _{2}$ remain finite as $\sin
\kappa _{2}$ vanishes.

The reverse case, when $\kappa _{1}$ is close to $0$ or $\pi $ but $\sin
\kappa _{2}\ncong 0$, can be obtained by interchanging the indices $%
1\leftrightarrow 2$ and $\nu \leftrightarrow \nu ^{-1}$.

This results shows that the dynamics of the spin angles is well described
even if one of $\sin \kappa _{i}$ evolves through zero.

\end{document}